\documentclass[a4paper,11pt]{article}
\usepackage{jinstpub} 
\usepackage{lineno}

\usepackage{lscape}
\usepackage{xcolor}

\usepackage{subcaption}
\usepackage{graphicx}

\title{\boldmath Design, fabrication and characterization of 8x9 n-type silicon pad array for sampling calorimetry}

\author[a]{Sawan}
\author[a]{G.~Tambave}
\author[b]{J.~L.~Bouly}
\author[b]{O.~Bourrion}
\author[c]{T.~Chujo}
\author[d]{A.~Das}
\author[e]{M.~Inaba}
\author[a]{V.~K.~S.~Kashyap}
\author[f]{C.~Krug}
\author[d]{R.~Laha}
\author[g]{C.~Loizides}
\author[a,1]{B.~Mohanty\note{Corresponding author.}}
\author[a]{M.M.~Mondal}
\author[b]{N.~Ponchant}
\author[a]{K.~P.~Sharma}
\author[a]{R.~Singh}
\author[b]{D.~Tourres}

\affiliation[a]{National Institute of Science Education and Research, Homi Bhabha National Institute, Jatni, India}
\affiliation[b]{Univ. Grenoble Alpes, CNRS, Grenoble INP, LPSC-IN2P3, 38000 Grenoble, France\\
and Institute of Engineering Univ. Grenoble Alpes}

\affiliation[c]{ Institute of Pure and Applied Sciences, University of Tsukuba, Tsukuba, Japan}
\affiliation[d]{Components SBU, Bharat Electronics Limited, Bangalore, India}
\affiliation[e]{Tsukuba University of Technology, Japan}
\affiliation[f]{Instituto de Física, Universidade Federal do Rio Grande do Sul, Porto Alegre, RS, Brazil}
\affiliation[g]{ORNL, Oak Ridge, USA}

\emailAdd{bedanga@niser.ac.in}

\abstract{This paper reports the development and testing of n-type silicon pad array detectors targeted for the Forward Calorimeter (FoCal) detector, which is an upgrade of the ALICE detector at CERN, scheduled for data taking in Run~4~(2029-2034). The FoCal detector includes hadronic and electromagnetic calorimeters, with the latter made of tungsten absorber layers and granular silicon pad arrays read out using the High Granularity Calorimeter Readout Chip~(HGCROC). This paper covers the Technology Computer-Aided Design (TCAD) simulations, the fabrication process, current versus voltage (IV) and capacitance versus voltage (CV) measurements, test results with a blue LED and $^{90}$Sr beta source, and neutron radiation hardness tests. IV measurements for the detector showed that 90\% of the pads had leakage current below 10~nA at full depletion voltage. Simulations predicted a breakdown voltage of 1000~V and practical tests confirmed stable operation up to 500~V without breakdown. CV measurements in the data and the simulations gave a full depletion voltage of around 50~V at a capacitance of 35~pF. LED tests verified that all detector pads responded correctly. Additionally, the 1$\times$1 cm$^2$ pads were also tested with the neutron radiations at a fluence of $5\times10^{13}$ 1~MeV~n$_{eq}$/cm$^2$.}

\keywords{Si microstrip and pad detectors, Detector design and construction technologies and materials, Calorimeters}

\begin{document}
\maketitle
\flushbottom

\section{Introduction}
\label{sec:intro}

Silicon detectors are widely used in particle physics experiments for their fast response, flexibility in design, and ability to withstand high radiation doses~\cite{ATLASpixeldetector, CMSbarrelPixelDetector}. The Forward Calorimeter (FoCal) which is a part of the ALICE detector upgrade for Run 4 at the Large Hadron Collider (LHC), utilizes silicon detectors for calorimetry and tracking in its electromagnetic component.\\
Starting in 2029, the ALICE experiment plans to install the FoCal detector 7 meters away from the ALICE interaction point. FoCal aims to measure small-$\it{x}$ (Bjorken scaling factor) gluon distributions via the measurements of direct photons in the forward pseudo-rapidity range of 3.4 < $\eta$ < 5.8. This will support ALICE in conducting inclusive and correlation measurements of photons, mesons, and jets to study the dynamics of the Quark-Gluon Plasma (QGP) at small-$\it{x}$ down to $10^{-6}$~\cite{focal_LOI}.
The ALICE FoCal consists of an Electromagnetic Calorimeter (FoCal-E) and a Hadronic Calorimeter (FoCal-H). The FoCal-E is made up of 20 alternating layers of silicon detectors and tungsten absorbers. Eighteen of these silicon layers are arrays of 8~$\times$~9 Si pads, each 1 cm$^2$, fabricated on 325 $\mu$m thick, 6-inch Si wafer. Additionally, two high-granularity CMOS pixel layers with an individual pixel size of 30~$\times$~30 $\mathrm{\mu m^{2}}$ are included for high position resolution.\\
Based on the substrate, silicon detectors are categorized into n-type and p-type detectors. A n-type detector has a n-type silicon substrate with a high concentration of electron donors (n+) on one side and electron acceptors (p+) on the other. Conversely, a p-type detector has a p-type substrate with p+ on one side and n+ on the other. This article focuses on the study of 8$\times$9 n-type silicon pad array detectors and their characterization in the laboratory, in addition and for comparison with the p-type pad arrays presented in the Technical Design Report (TDR) \cite{focaltdr}. These detectors are fabricated on 6-inch, 325~$\muup$m~$\pm$~20~$\muup$m thick, high-resistivity ($\sim$~7~k$\Omega\thinspace$cm) silicon wafers, with each array containing 72 single pads of area 1~$\times$~1 cm$^{2}$ per pad. Similar silicon detector designs and their use in electromagnetic calorimeters have been previously reported ~\cite{Clice_ILC_1, Clice_ILC_2, Clice_ILC_3, Tsukuba_paper, sanjib_paper}. The large-area n-type (8$\times$9) detectors discussed here are being fabricated for the first time in India. The size requirements of the silicon pad arrays are based on the HGCROCv2 readout chip~\cite{HGCROCv2_paper}, which has 72 readout channels matching the number of Si pads and therefore can be directly attached to the Si detector, reducing the overall detector size. This paper will discuss the detector design, Technology Computer-Aided Design (TCAD) device simulation, fabrication process, and electrical and performance tests using a light-emitting diode (LED) and a $^{90}$Sr electron source. Additionally, it will report on radiation hardness studies conducted using a fast neutron beam at RANS, RIKEN, Japan. For the response of the detector to high-energy pion and electron beams refer to paper \cite{testbeampaper}.

\section{Silicon pad array design}
\begin{figure}[t]
\centering
\includegraphics[width=1.0\textwidth]{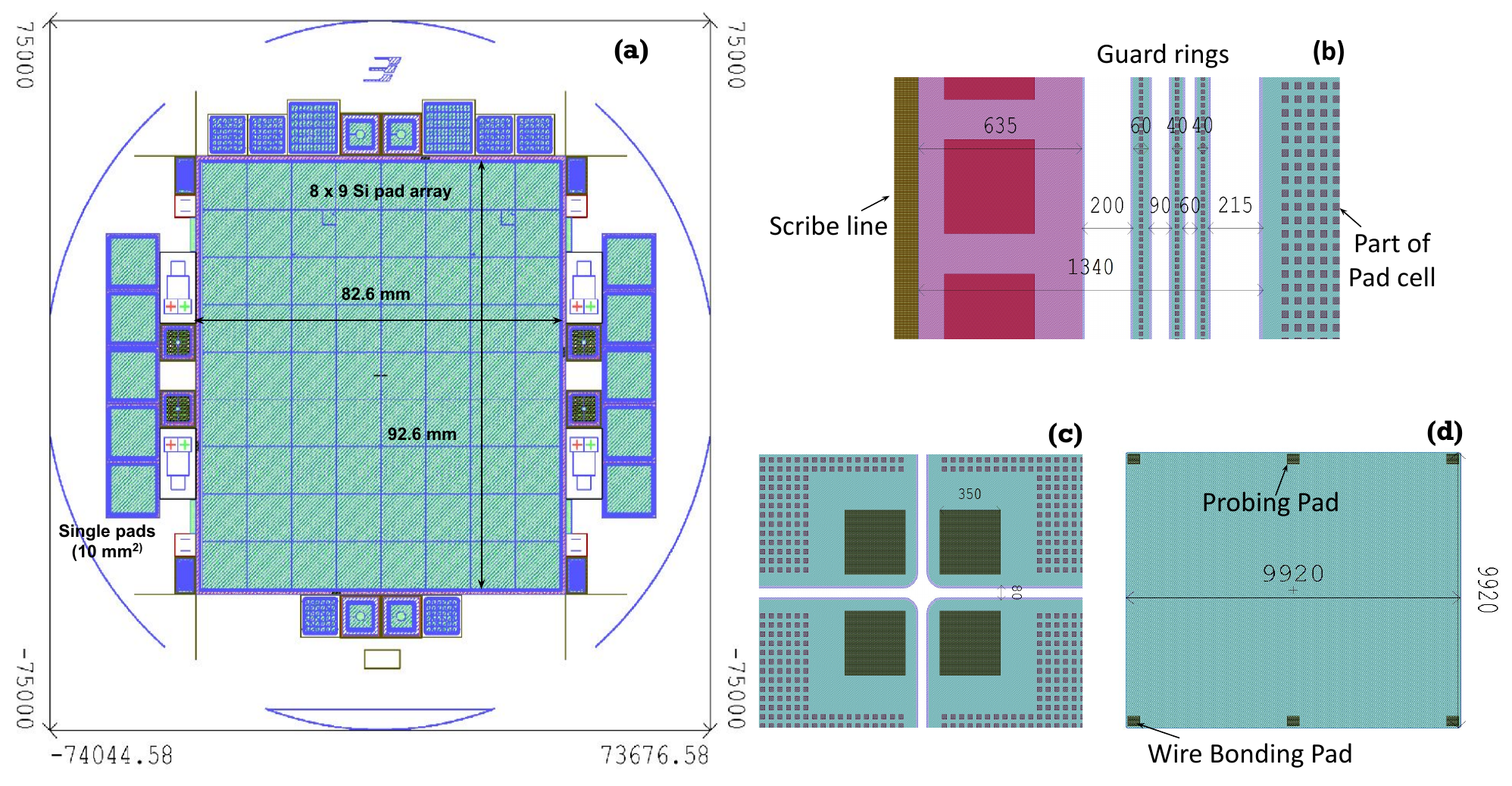}
\caption{Design of the 8~$\times$~9 Si pad array: (a) Wafer-level mask design layout showing main pad array in the center and small pad cells in the periphery, (b) Guard ring design showing the three guard rings, inter-guard ring distances, and scribe line, (c) Zoomed-in view of the layout showing inter pad distance of around 80~$\muup$m, (d) Layout of a single pad cell showing positions of the probing and wire bonding pads, respectively. Dimensions are in~$\muup$m.
\label{fig:padarraydesign}}
\end{figure}
The n-type Si pad array design consists of an 8~$\times$~9 array of pad cells, with each cell measuring 1~$\times$~1~cm$^{2}$. Figure 1(a) shows the full wafer-level design layout, with the main die (82.6~mm~$\times$~92.6~mm) in the center and test samples of pad cells around the wafer's periphery. Three guard rings are located at the outer edge of the pad array, as seen in Figure 1(b), which also shows part of the pads and the scribe line, with dimensions in micrometers. The pad cell is approximately 1340 $\muup$m away from the scribe line. Figure 1(c) illustrates that the pads are separated by a small distance of 80~$\muup$m to reduce the cross-talk in the detector and also ensure a minimal dead area in the array. Each pad cell has six contact points, shown as dark green squares (350~$\muup$m$^{2}$) in Figure 1(d). The wire bonding pads at the corners connect the pad cells to readout electronics, while the probing pads are used for electrical testing of the detector after fabrication.
\section{TCAD device simulation}
\begin{figure}[t]
\centering
\includegraphics[width=.6\textwidth]{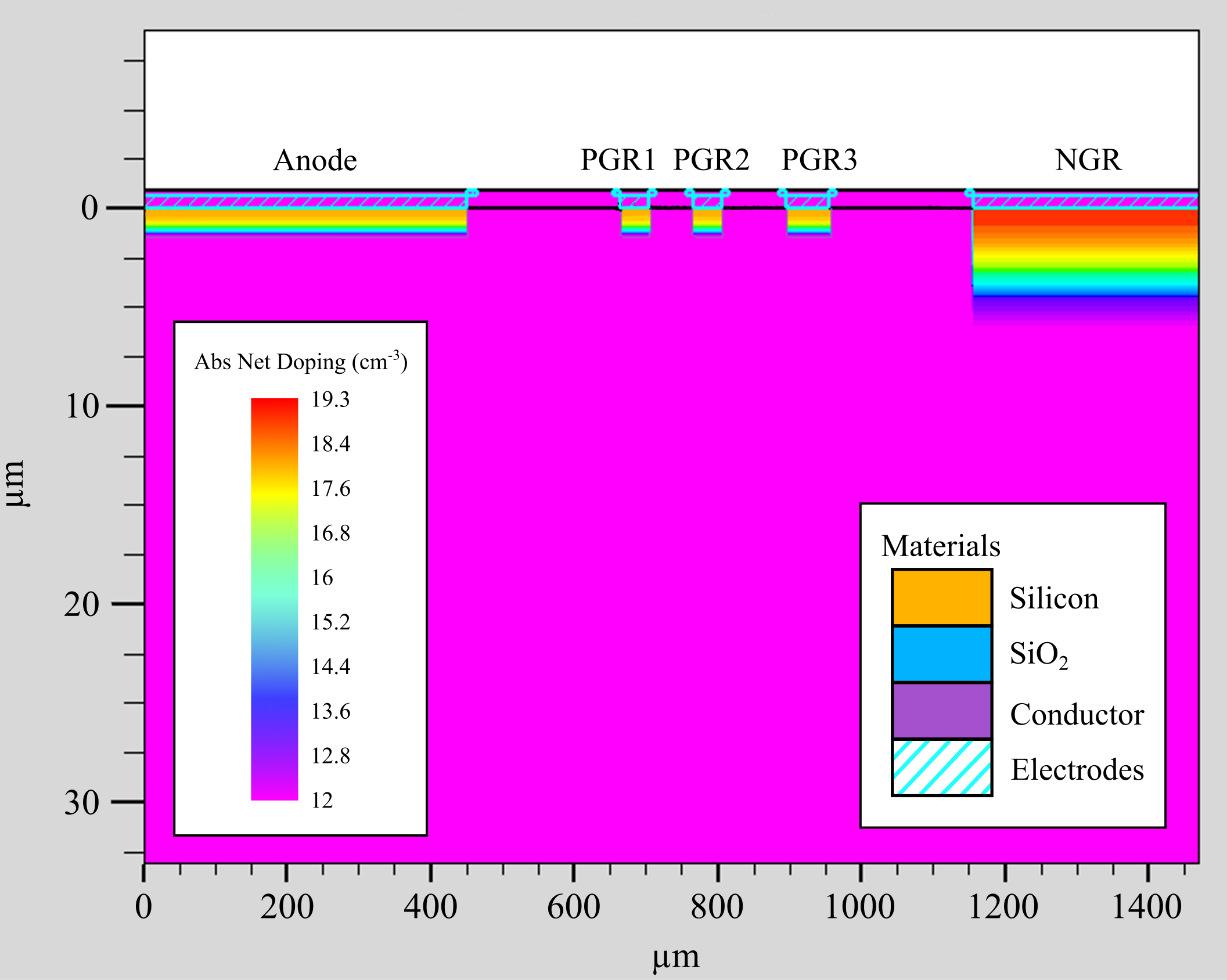}
\caption{The design and doping profile of the 8×9 Si pad array: The layout includes the anode, three p-type guard rings (PGR1, PGR2, PGR3), and a n-type guard ring (NGR). The vertical axis represents depth in microns, while the horizontal axis shows the position in microns. The color gradient indicates the absolute net doping concentration in cm$^{-3}$ whose values are listed in the left legend. The materials are labeled with colors: silicon (orange), SiO$_{2}$ (blue), conductor (purple), and electrodes (striped blue).\label{fig:TCad_Drawing}}
\end{figure}

The detector fabrication process and its electrical characteristics are simulated using the Athena and Atlas modules of Silvaco Technology Computer-Aided Design (TCAD) software \cite{TCAD_simulations}. For this simulation, a Passivated Implanted Planar Silicon (PIPS) type standard P+/N-/N+ vertical stack was used as shown in Figure~\ref{fig:TCad_Drawing}. To reduce mesh size and computational time, only the edge of the end pad cell and the guard ring structures (three p-type guard rings and one n-type guard ring) were simulated. The color legend on the left indicates the doping concentration, where the top of the n-type substrate wafer was doped with boron impurities at 80 keV energy to form the p+ regions (anode and p-type guard rings) and with the n-type phosphorus impurities to form the n+ regions (n-type guard ring (NGR)). The NGR is introduced to prevent the depletion region from reaching the detector edge, which could cause premature breakdown.
The simulated junction depth exceeds 1 micron for both the active area and the p-type guard rings. The simulation also accounts for the effects of metal overhang and field oxide, ensuring a comprehensive analysis of the electrical characteristics of the detector.
The various parameters such as implant dose, drive-in time and temperature, guard ring distance, metal overhang, oxide thickness, etc. were optimized for the highest possible breakdown voltage of the detector.

\begin{figure}[t]
\centering
\includegraphics[width=.4\textwidth]{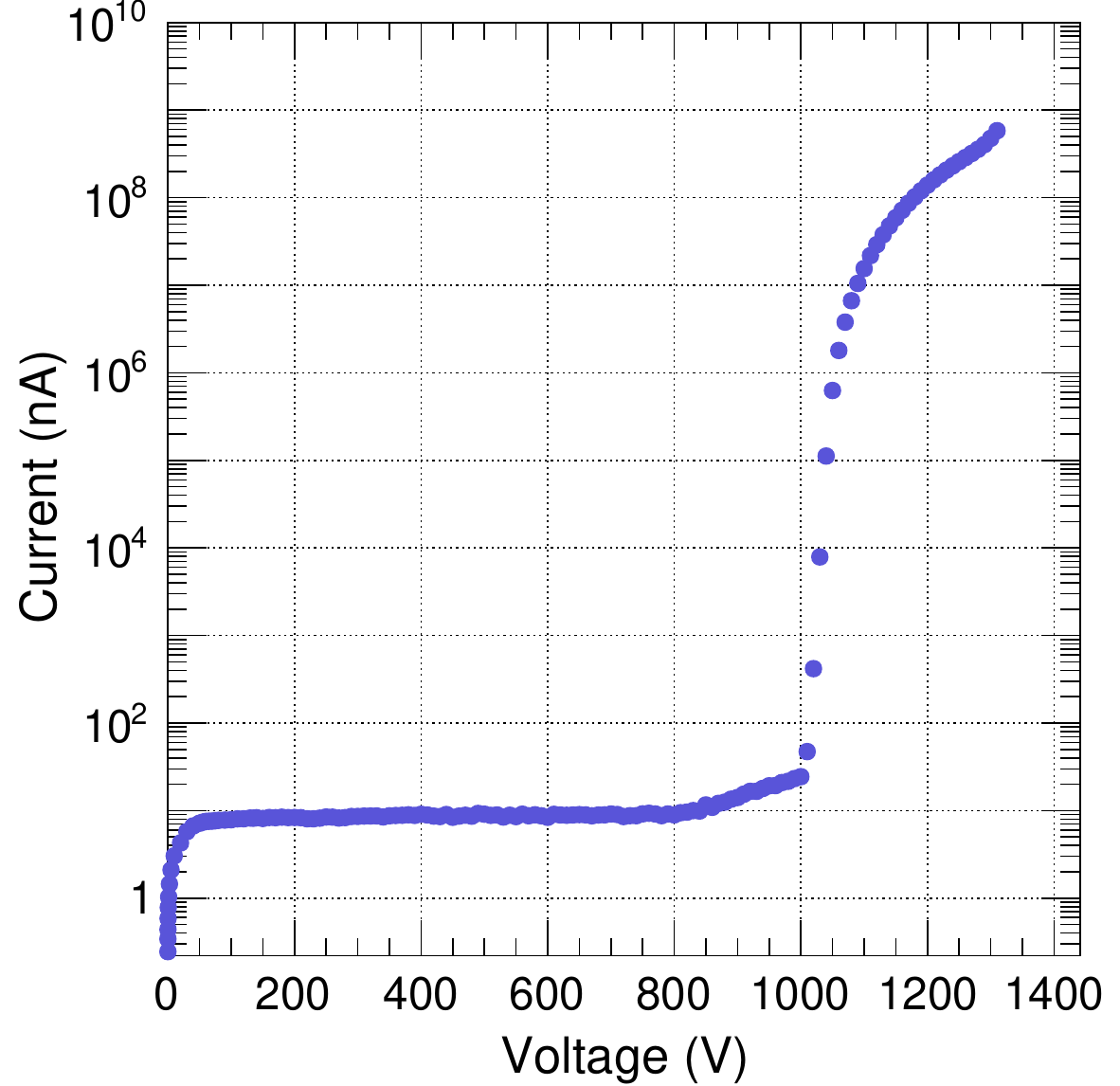}
\qquad
\includegraphics[width=.4\textwidth]{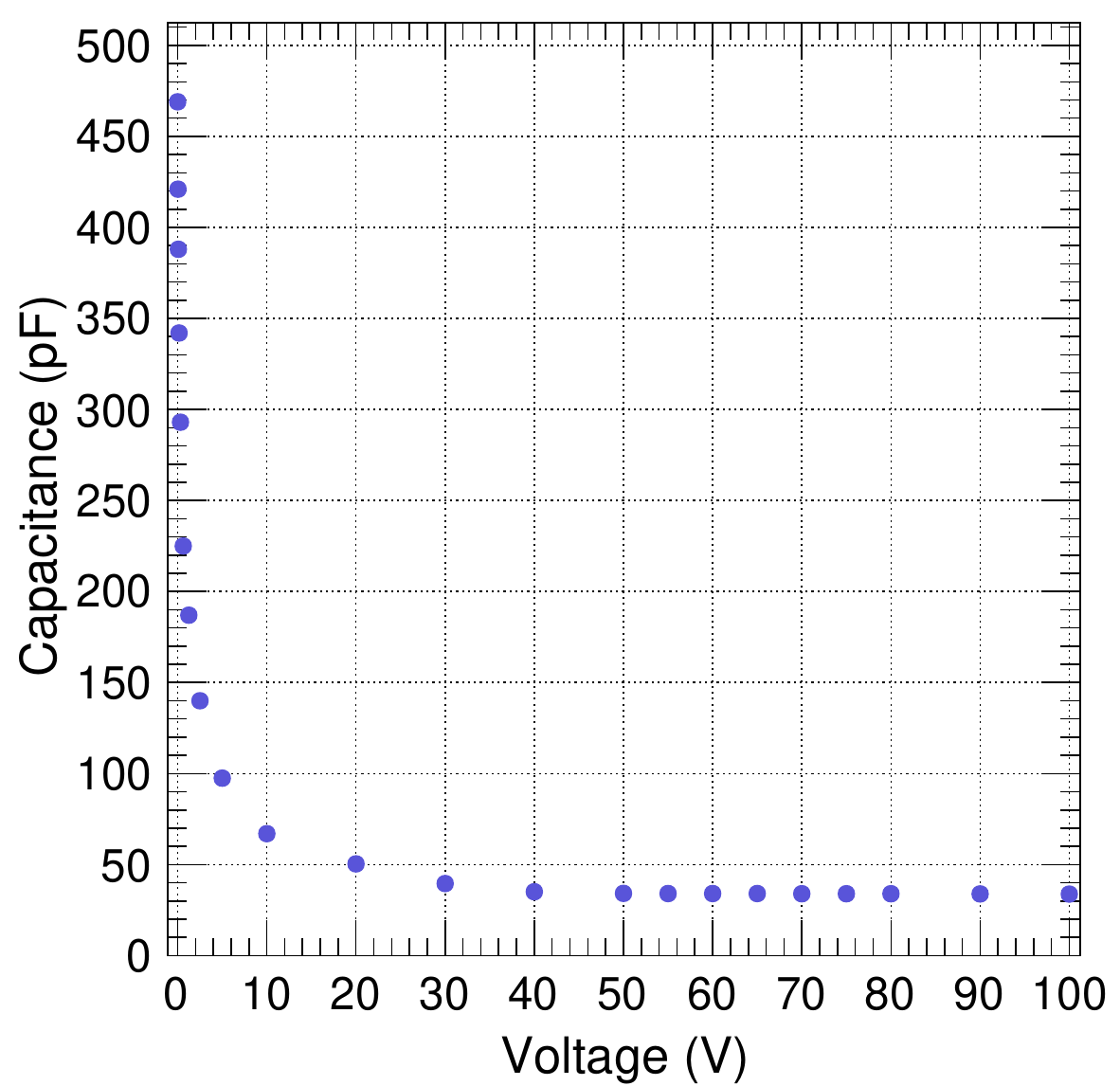}
\caption{Simulated IV (Left) and CV (Right) curves of 1$\times$1 cm$^2$ pad cell.\label{fig:SimulatedIV-CV}}
\end{figure}

In the simulation, a part of the pad cell was considered where the anode and cathode in the design were kept at the negative and positive potential to simulate the leakage current as well as the device capacitance as a function of reverse bias voltage. The simulated IV and CV plots are shown in Figure~\ref{fig:SimulatedIV-CV}. The IV characteristics indicate a breakdown voltage exceeding 1000~V, while the CV characteristics suggest a full depletion voltage between 40-50~V.

The mask layout (see Figure~\ref{fig:padarraydesign}~(a)) was created using L-edit layout editor software, incorporating the optimized simulation results. The PCB layout was superimposed on the final photomask Graphic Data System (GDS) to ensure proper alignment of the wire bond pads. Before finalizing the detector design layout, an alignment test was conducted in the software. This test confirmed that the wire bonding pads on the Si detector were perfectly aligned with the center of the 2.1~mm circular opening on the Front-End-Electronics (FEE) board. The FEE board is a 10-layer printed circuit board (PCB) hosting the currently available version 2 of HGCROC chip~\cite{bourrion2023prototype}.


\section{Si pad array (8~$\times$~9) fabrication}\label{sec:fabrication}
\begin{figure}[t]
\centering
\includegraphics[width=.6\textwidth]{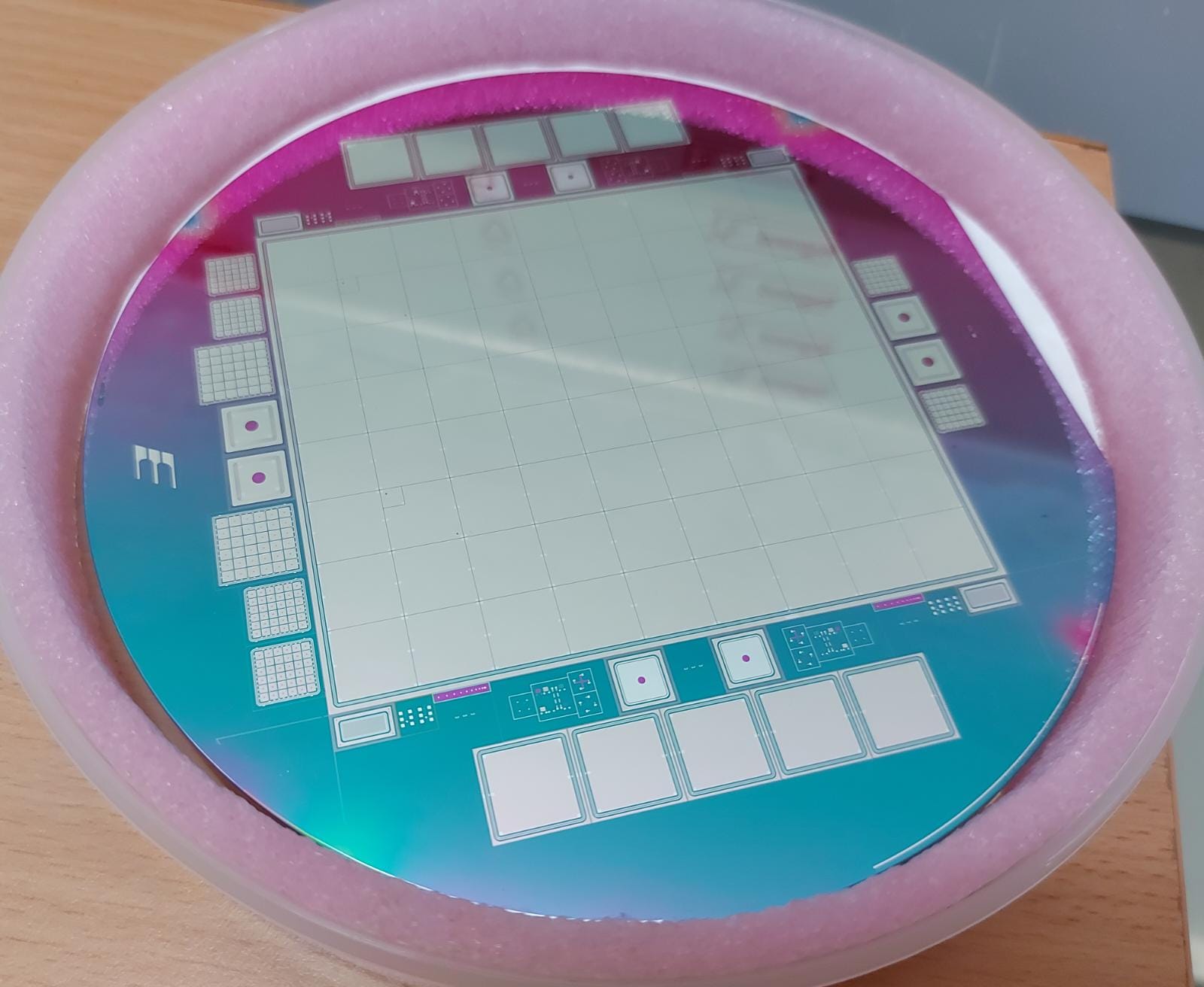}
\caption{Photograph of fabricated Si pad array on a 6-inch n-type Si wafer.\label{fig:wafer-photo-probeCard}}
\end{figure}

Based on the TCAD device simulations and the design discussed in the previous section, several samples of the n-type Si pad array detector were fabricated. The target parameters for the device fabrication were: reverse saturation current or leakage current less than 10~nA, detector capacitance of about 35~pF at a full depletion voltage (FDV) of around 50~V, and a breakdown voltage of 500~V or more per pad cell. 
To achieve these parameters, the detectors were fabricated using Si wafers and photomasks with the specifications listed in Table~\ref{Siwafer-table} and Table~\ref{photoMask-table}. The fabrication process involved using a five-layered photomask and a negative photoresist. Standard bipolar fabrication technology was employed at the Bharat  Electronics Limited~(BEL) wafer fab in Bangalore, India \cite{bel}.
The fabrication process included steps such as thermal oxidation, photolithography, ion implantation, wet and dry etching, diffusion, metallization, and protective layer deposition. The cleaning and etching steps were optimized through trials to minimize contamination and achieve the lowest possible leakage current at full depletion voltage. Additionally, the diffusion and oxidation steps were optimized to maintain the uniformity of the wafer within~10\%. After these steps a 6~$\mu$m aluminum layer was deposited on the backside (n+) of the detector. A photograph of the finished wafer is shown in Figure~\ref{fig:wafer-photo-probeCard}.
\begin{table}[t]
\begin{center}
\caption{Specification of Si wafers used for device fabrication.}
\label{Siwafer-table}
\begin{tabular}{||l||p{0.3\linewidth}||}
 \hline
 \textbf{Parameter} & \textbf{Description} \\  
 \hline\hline
 Type & n-type prime, Single crystal Si \\ 
 \hline
 Growth & Float Zone (FZ) \\
 \hline
 Diameter & 150 $\pm$ 0.5 mm \\
 \hline
 Orientation & $\langle 1\ 0\ 0 \rangle$ \\
 \hline
  Thickness & 325 $\pm$ 20 $\muup$m \\
 \hline
  Substrate resistivity & $\sim$ 7~k$\Omega\thinspace$cm  \\
  \hline
  Dopant & Phosphorus, n-type \\
  \hline
  Total Thickness Variation (TTV) & 10 $\muup$m \\
  \hline
  Maximum Oxygen and Carbon Conc. & $10^{16}$/cm$^3$ \\
  \hline
  Minority carrier recombination lifetime & 1 ms \\
  \hline
\end{tabular}
\end{center}
\end{table}
\begin{table}[!t]
\begin{center}
\caption{Specification of photo-mask used during device fabrication.}
\label{photoMask-table}
\begin{tabular}{||l||p{0.4\linewidth}||}
 \hline
 \textbf{Parameter} & \textbf{Description} \\  
 \hline\hline
 Material & Soda lime glass / Anti-reflective chrome \\ 
 \hline
 Dimensions & $7^{\prime\prime}\times 7^{\prime\prime}\times 0.12^{\prime\prime}$ \\
 \hline
 Pattern generation  & Direct write \\
 \hline
 Max. defect size & 0.5 $\muup$m \\
 \hline
 Max. defect density & 0 \\
 \hline
 Min. size of layout elements in layer & 10 $\muup$m $\pm$ 1~$\muup$m \\
 \hline
 Writing grids & 0.1~$\muup$m \\
 \hline
 Mask order &  7$^{\prime\prime}$/Bright Field (BF)/Chrome Down (CD)/Right Reading (RR) (for 4 layers) and 7$^{\prime\prime}$/Dark Field (DF)/CD/RR (for 1 layer)\\
 \hline
 Max. misalignment between layers & 0.5 $\muup$m \\
 \hline
\end{tabular}
\end{center}
\end{table}
\section{Detector integration with FEE}
\begin{figure}[t]
\centering
\includegraphics[width=1.0\textwidth]{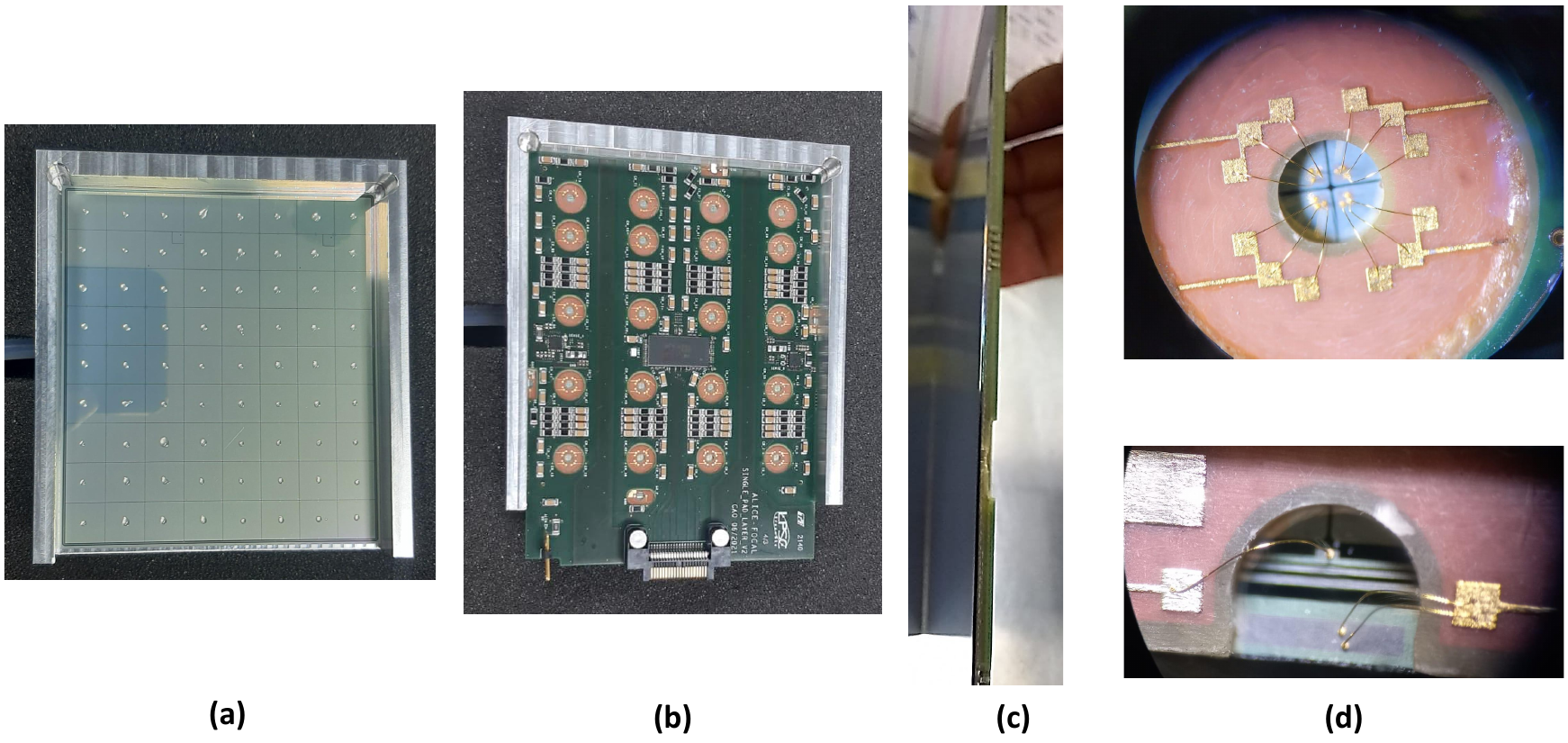}
\caption{The photographs show the packing procedure of Si detector (die): (a) The die (82.6~mm$\times$~92.6~mm) is held in place inside the Al jig with glue dots added at the center of each pad, (b) FEE board (82.6~mm$\times$~100.6~mm) is placed on top of the detector and left for 12~hours for the glue to settle, (c) Side view of FEE board glued to the detector, and (d) 25~$\mu$m thick gold wire bonds connected to the corners of four pads (top), and gold wire bonds connecting the two floating guard rings and one guard ring to the ground (bottom).
\label{fig:PadArrayAssembly}}
\end{figure}
After completion of device fabrication, wafer-level electrical tests (IV/CV) were carried out. The good wafers that met the design criteria mentioned in section~\ref{sec:fabrication}, were selected for dicing the pad array (die). The procedure followed in attaching the die to the FEE board is depicted in Figure~\ref{fig:PadArrayAssembly}. An aluminum jig of the same dimensions as the die was made with a provision to apply suction from the bottom so that the die would be temporally fixed in the jig, then manually Sader epoxy glue dots were placed at the center of each pad cell as shown in Figure~\ref{fig:PadArrayAssembly}~(a). The FEE board was gently placed on top of the die as shown in Figure~\ref{fig:PadArrayAssembly}~(b) and rested in place overnight so that, the glue could settle and make a robust packaged detector. The side view of the packaged detector is shown in the photograph of Figure~\ref{fig:padarraydesign} (c). A 25~$\muup$m thick gold wire bonding was carried out on the packaged detector. The upper panel in Figure~\ref{fig:padarraydesign} (d) shows a zoomed picture of one of the openings on the FEE board where gold wire bonds from the FEE board to the bonding pads on the four pad cells are shown. Each pad cell is connected with the three wire bonds to reduce the inductive reactance of wires. There are in total 20 circular openings on the FEE board, where the four openings in the top row connect to two bonding pads each, while the remaining 16~openings connect to four bonding pads each. This way all the the 72~pads of the detector array are connected to the FEE board. The bottom panel in Figure~\ref{fig:padarraydesign} (d) shows the wire bonds to the innermost and outermost guard rings. The inner guard ring is kept at the same voltage as the detector, the middle two guard rings are kept floating and the outermost guard ring is connected to the ground. After wire bonding, the glob top was done using transparent glue (Dow Sylgard 186 silicone elastomer) to protect the wire bonds. The details about the FEE are discussed in the following section~\ref{subsec:LED-test}.
\section{Test results and discussion}
 In this section, various tests of the fabricated and packaged Si pad array detectors are reported. Test done using bare detectors include electrical characteristics: IV (leakage current versus voltage) and CV (junction capacitance versus voltage), breakdown voltage test, and effect of temperature on the leakage current. For packaged detectors, the tests include the data acquisition optimization using a blue LED and the detector response to the $^{90}$Sr $\beta^{-}$ source which emits electrons up~to 2.2~MeV energy. In addition, radiation hardness studies of the detector sample irradiated with a fast neutron fluence of $5\times 10^{13}$~1~MeV~$\mathrm{n_{eq}}$/cm$^{2}$ are also reported. 

\subsection{Electrical tests (IV and CV)}
\begin{figure}[t]
\centering
\includegraphics[width=0.58\textwidth]{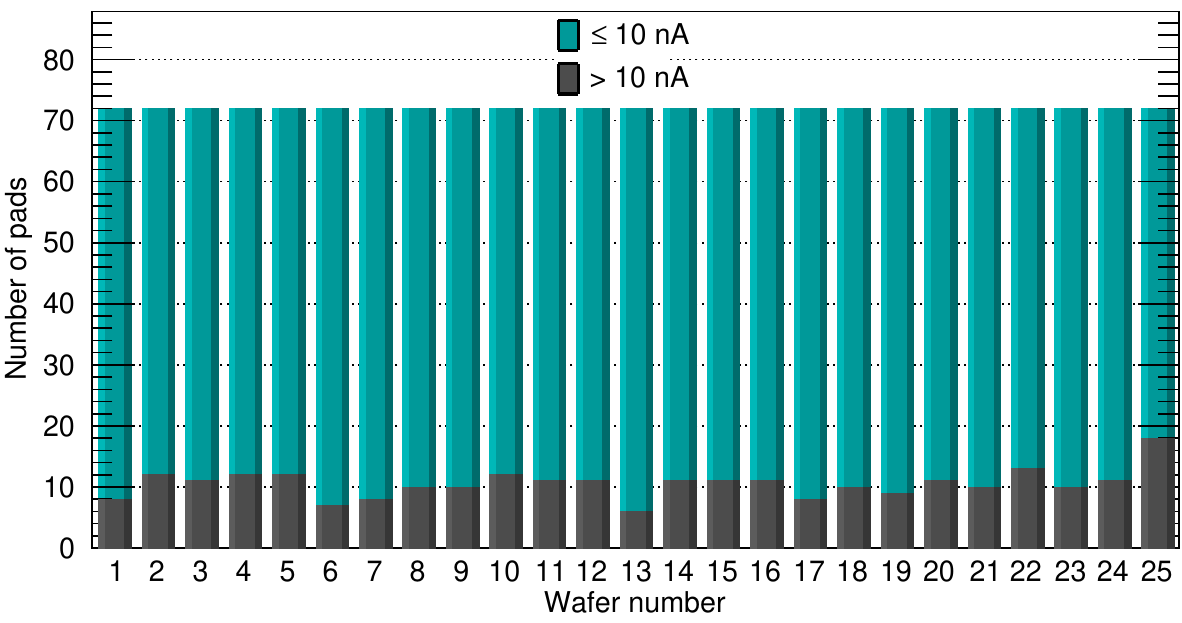}
\includegraphics[width=0.4\textwidth]{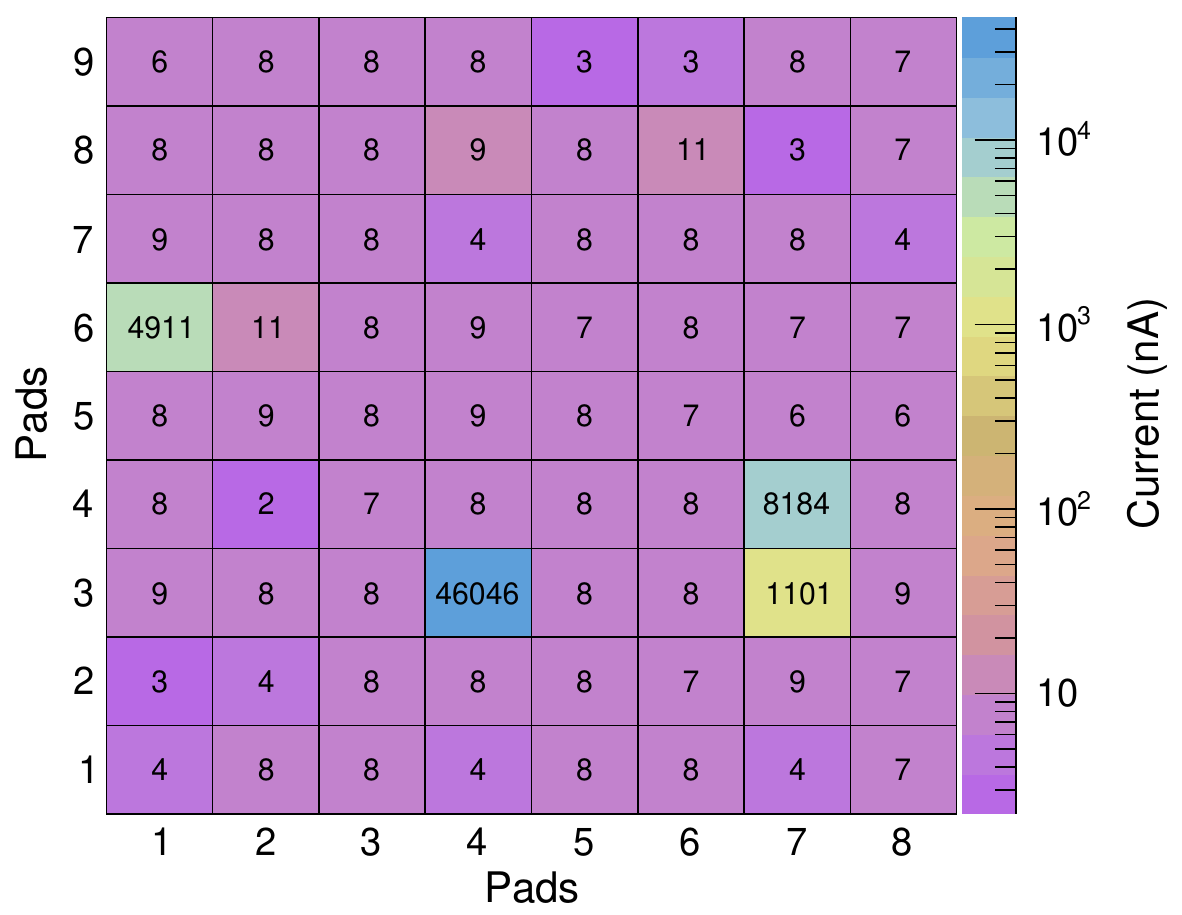}
\caption{(Left) The number of pad cells in a wafer with leakage current less than or equal to 10~nA (teal) and greater than 10~nA (grey) plotted as a function of wafer number. (Right) Leakage current map for a single pad array at a reverse bias voltage of 50~V.\label{fig:IV_allWafers}}
\end{figure}

\begin{figure}[t]
\centering
\includegraphics[width=0.4\textwidth]{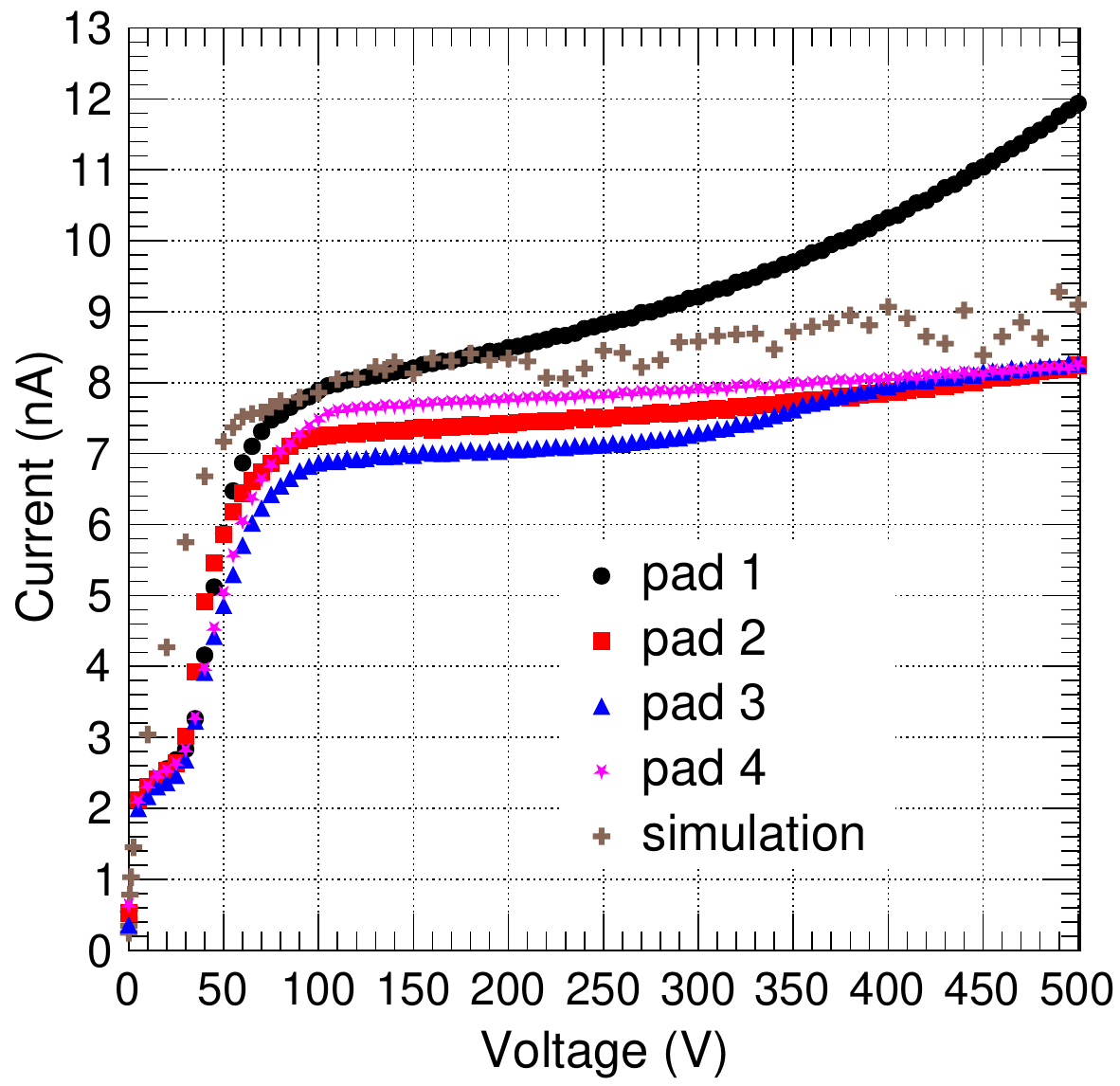}
\includegraphics[width=0.4\textwidth]{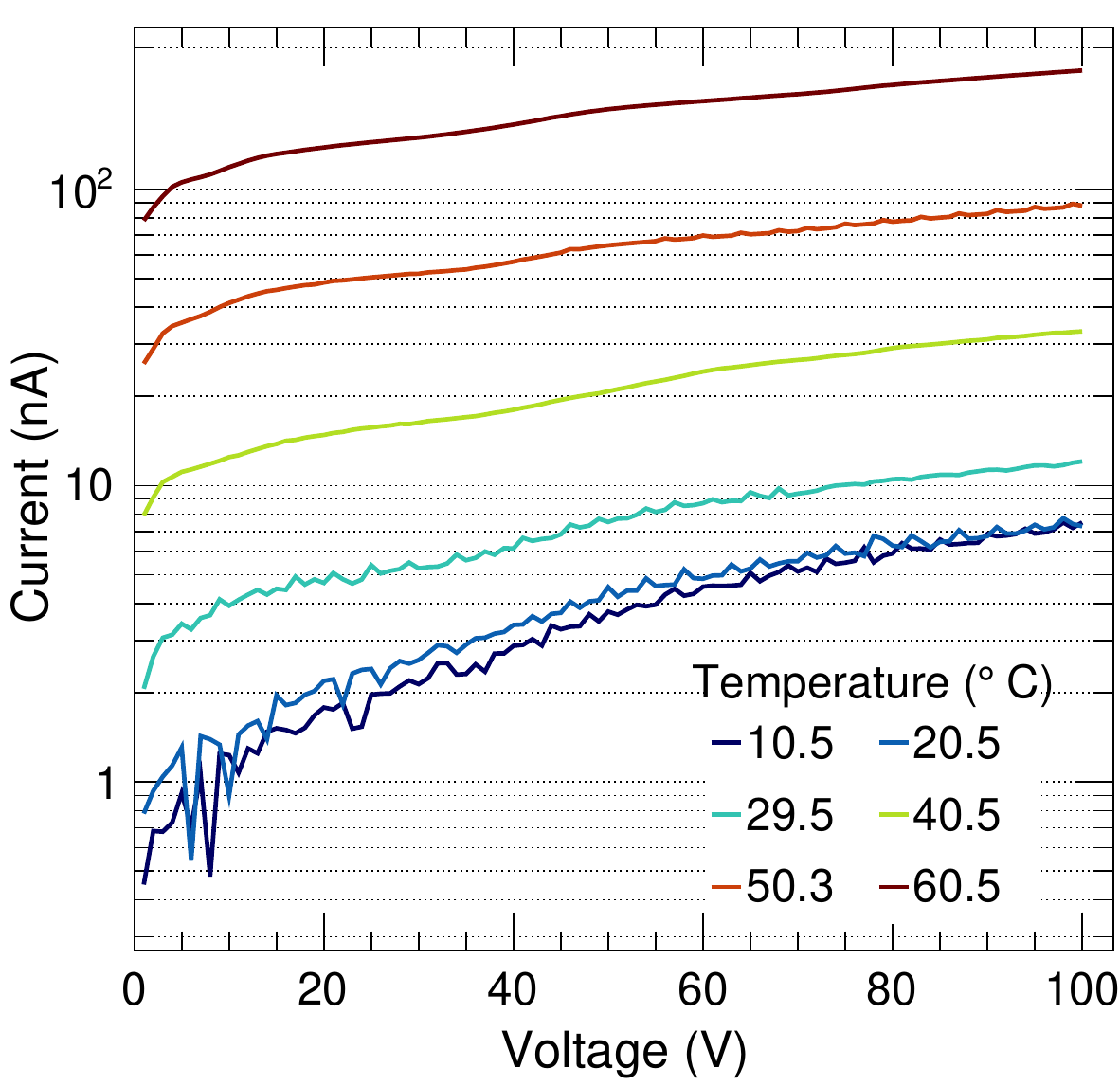}
\caption{(Left) Leakage current as a function of applied reverse bias voltage for a few randomly selected pads in the detector, shown along with the simulation result. (Right) Leakage current as a function of reverse bias voltage at various temperatures.\label{fig:IV_BR-temp}}
\end{figure}

This section reports the IV and CV tests of Si pad arrays carried out using the Keithley 4200A-SCS parameter Analyzer. The IV test was performed at the die level using a probe station. During the test, all neighboring pad cells were kept at the same potential as the pad cell under test (PCUT). This procedure was followed manually for each pad cell to acquire leakage current data at a fixed reverse bias voltage of 50~V. The leakage current for the best 25~Si pad arrays is shown in Figure~\ref{fig:IV_allWafers} (Left). About 90~\% of the pad cells have leakage current less than 10~nA/cm$^2$. While a lower leakage current is generally preferable, achieving an extremely low current over the large detector area was challenging, so the design aimed for a leakage current of less than 10~nA/cm$^2$. Figure~\ref{fig:IV_allWafers} (right) shows an example of a current-voltage map of a single pad array, where the leakage current in most pads stays below the 10~nA/cm$^2$ target.
To test the higher voltage tolerance of the detector beyond the full depletion voltage (FDV), four random pads on a detector array were probed (Figure~\ref{fig:IV_BR-temp} left). The results show the detectors can operate safely up to 500~V and might handle even higher voltages as expected from the simulations. However, for safety, the reverse bias voltage was not increased further. The measured current-voltage data is in agreement with the simulations.

Leakage current in semiconductor detectors is sensitive to temperature due to the generation of thermally induced electron-hole pairs~\cite{leobook1994}. To test the effect of temperature on the leakage current of the detector, a single pad cell was placed on a peltier module. Applying voltage across the peltier module causes one side to heat up while the other cools down. This setup allowed for measuring leakage current at reverse bias voltages up to 100~V, across temperatures ranging from 10.5~\textdegree C to 60.5~\textdegree C with interval of 10~\textdegree C as shown in Figure~\ref{fig:IV_BR-temp}. As expected the leakage current varies inversely as a function of temperature. The optimal operating condition is around 20~\textdegree C where the leakge current is well below 10~nA at FDV.
\begin{figure}[t]
\centering
\includegraphics[width=0.4\textwidth]{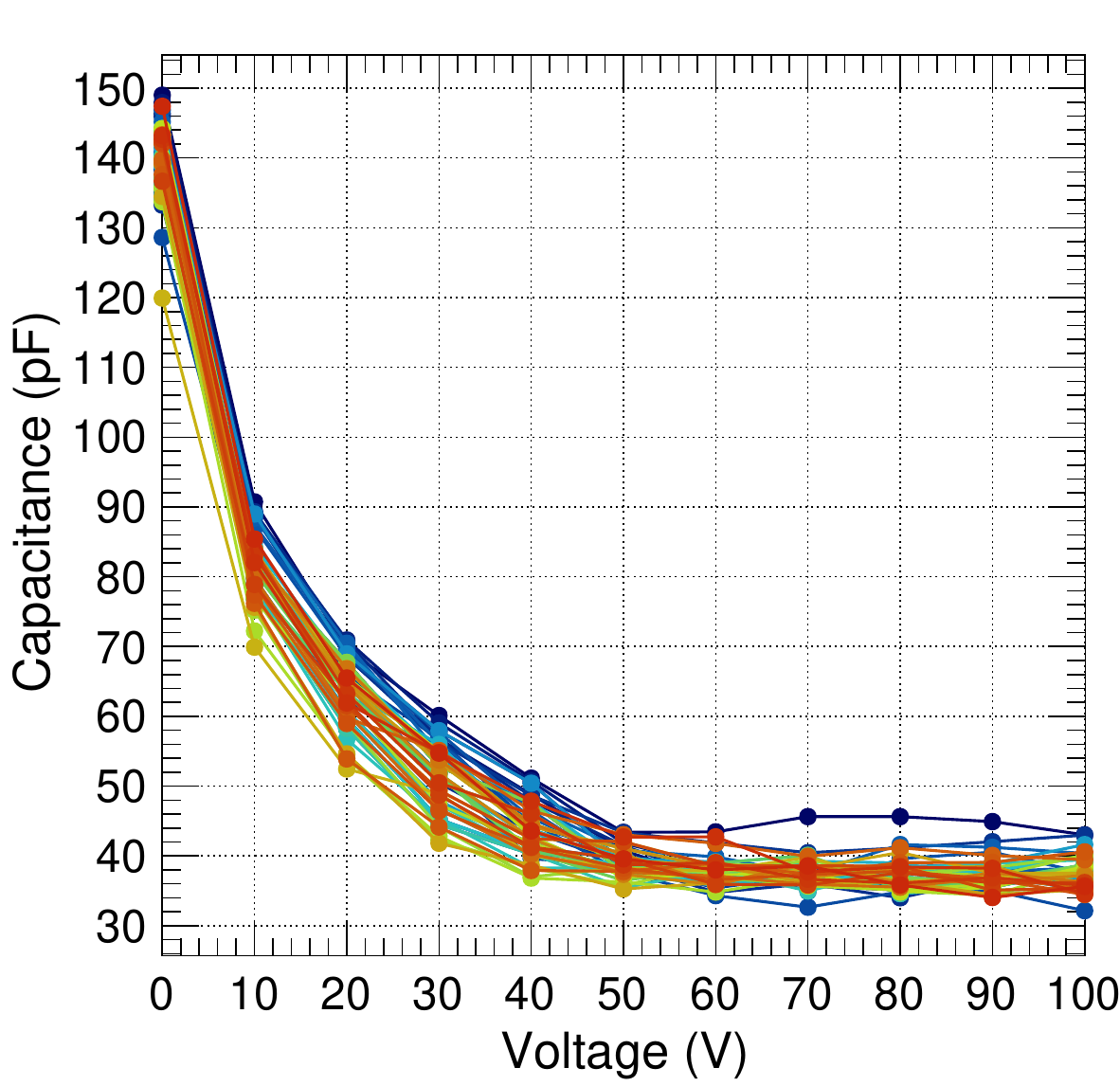}
\includegraphics[width=0.4\textwidth]{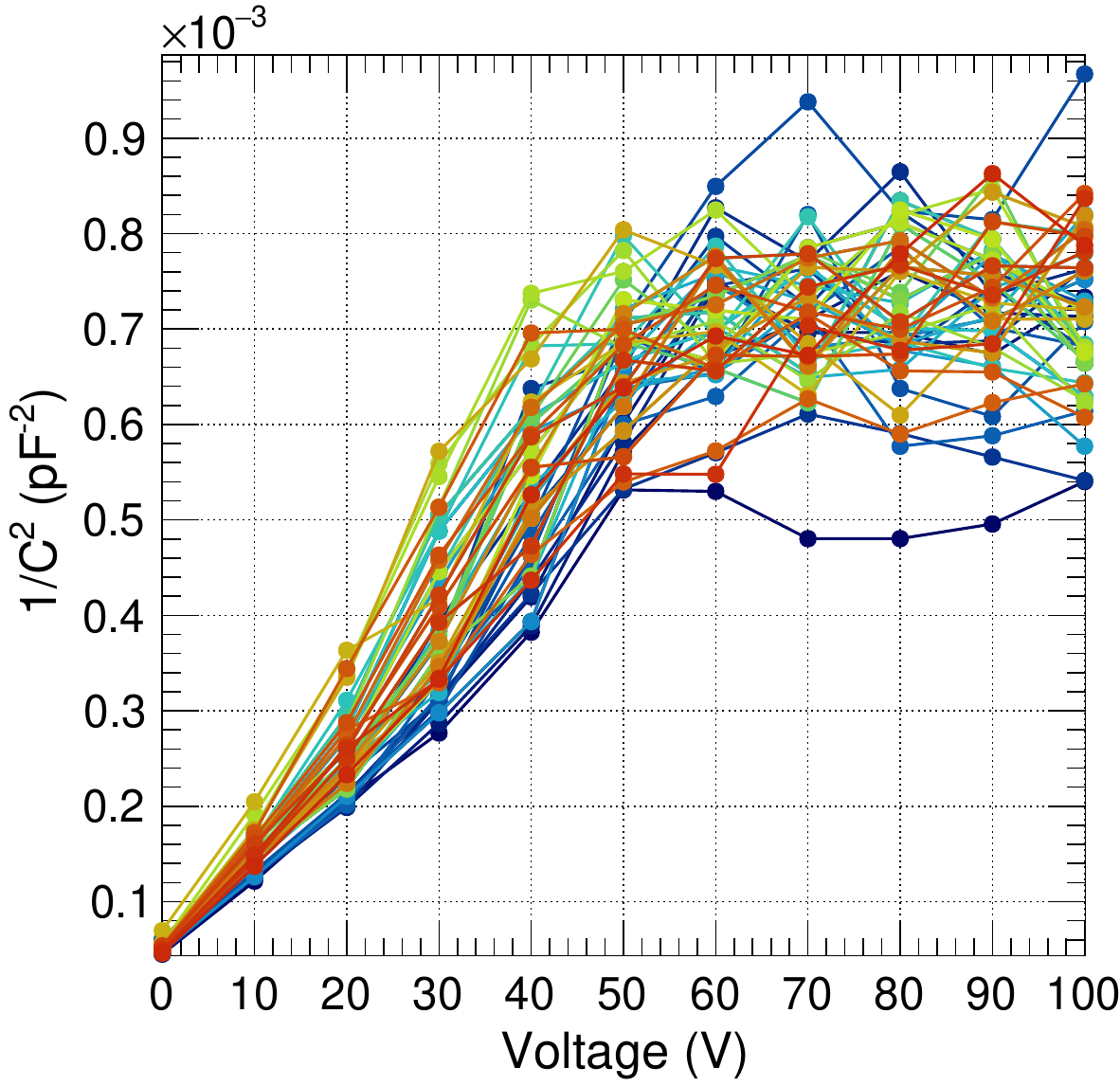}
\caption{(Left) Detector capacitance as a function of applied reverse bias voltage. (Right) The inverse square of capacitance as a function of applied voltage. \label{fig:measured-CV}}
\end{figure}

In addition to the IV measurements, CV measurements were performed to determine the full depletion voltage (FDV) of the detector and compare it with simulations. The CV characteristics for one of the detector array samples are shown in Figure~\ref{fig:measured-CV} (Left). The pads have a capacitance between 32~pF and 46~pF above 50~V, which is in accordance with the simulations. To obtain the FDV, the inverse of the capacitance squared ($1/C^{2}$) is plotted against the applied voltage for all 72 pads in Figure~\ref{fig:measured-CV} (Right). The voltage at which the $1/C^{2}$ values saturate is considered the FDV, which is around 50 V in this case. Therefore, the operating voltage of the detector would be slightly above the FDV, around 60 V, to ensure stable conditions.

\begin{figure}[t]
\centering
\includegraphics[width=0.7\textwidth]{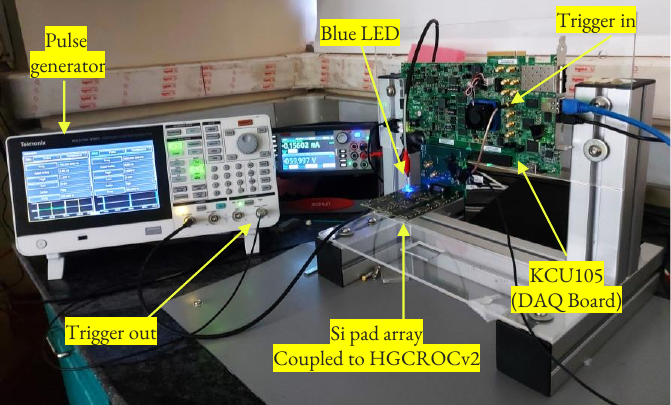}
\caption{The LED test setup used to obtain the trigger parameters for the configuration of HGCROCv2 chip: The setup includes a blue LED, a pulse generator, a Si pad array coupled to the HGCROCv2 chip, and an Xilinx AMD Kintex UltraScale FPGA KCU105 Evaluation Kit DAQ board. The blue LED illuminates the Si pad array, with the pulse generator driving the LED and controlling the trigger signals passed to the DAQ board. \label{fig:LED_pulser_photo}}
\end{figure}

\subsection{Detector test with LED}\label{subsec:LED-test}
\begin{figure}[t]
\centering
\includegraphics[width=.4\textwidth]{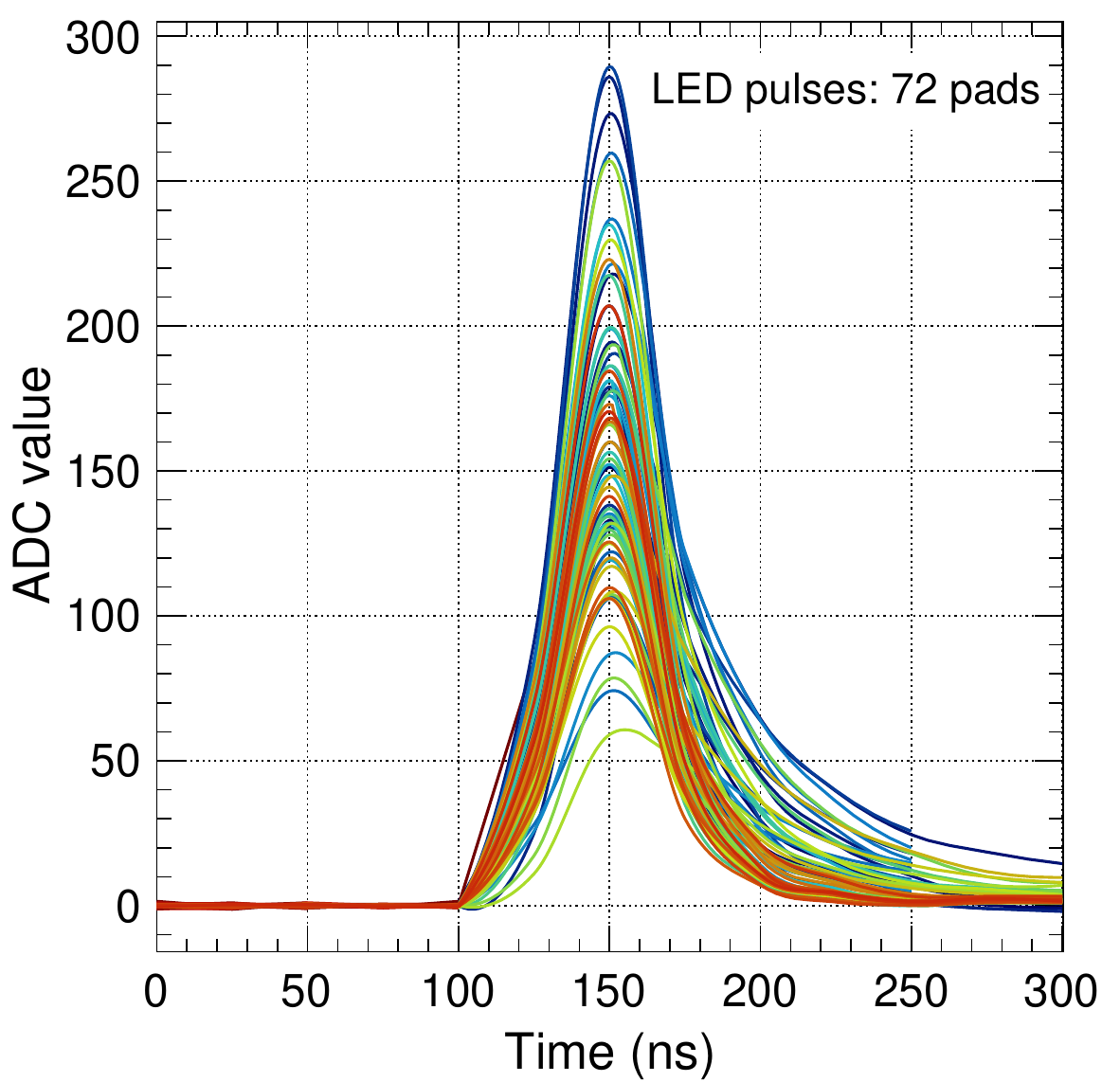}
\qquad
\includegraphics[width=.4\textwidth]{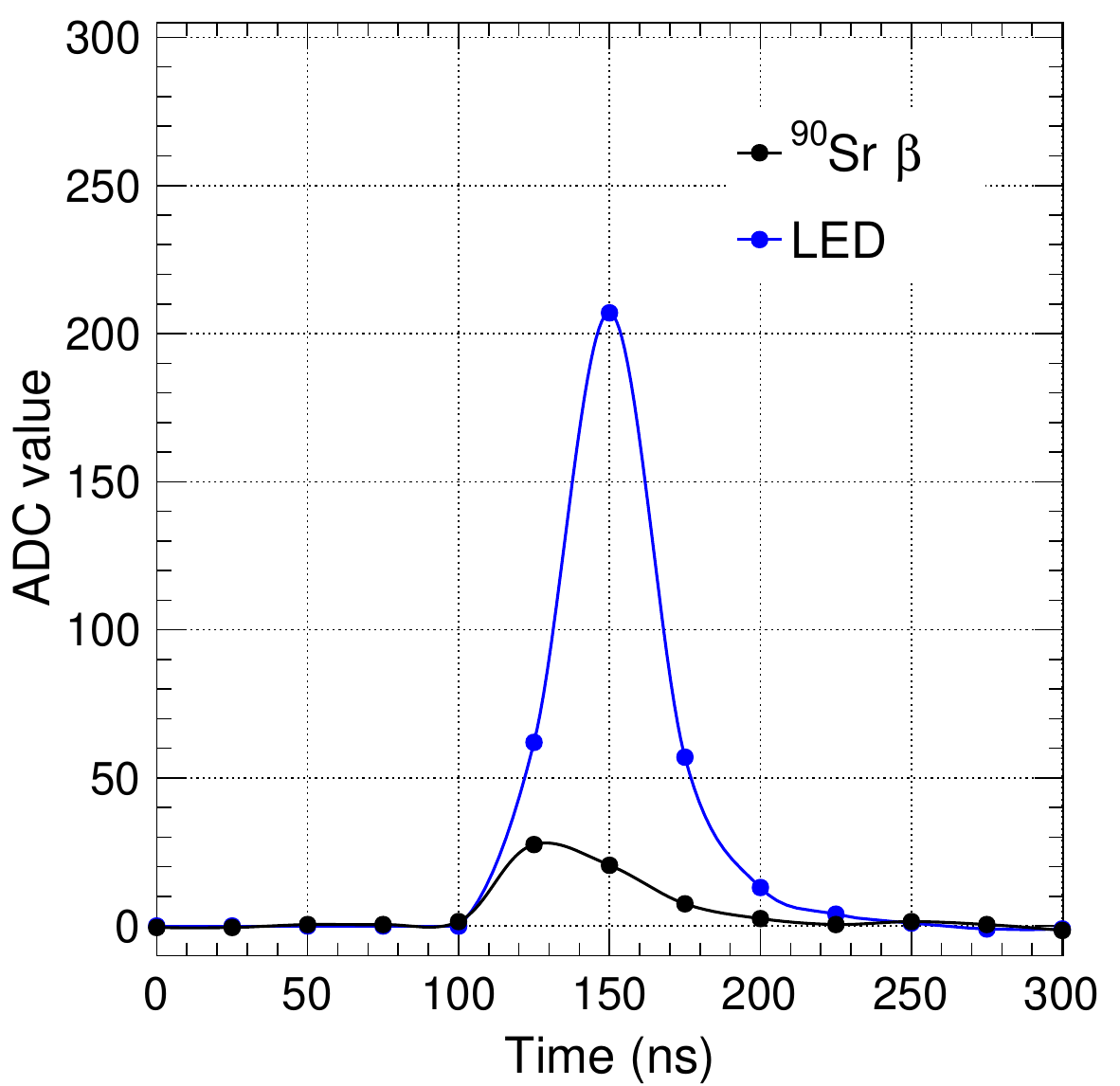}
\caption{This figure shows signal pulses measured in ADC counts (y-axis) and time in ns (x-axis). The left panel confirms the functionality of all 72 detector pads by showing their LED light pulses. The right panel compares the signal strength from $^{90}$Sr electrons with the blue LED light signal.\label{fig:sr90-LED-pulses}.} 
\end{figure}

\begin{figure}[t]
\centering
\includegraphics[width=.9\textwidth]{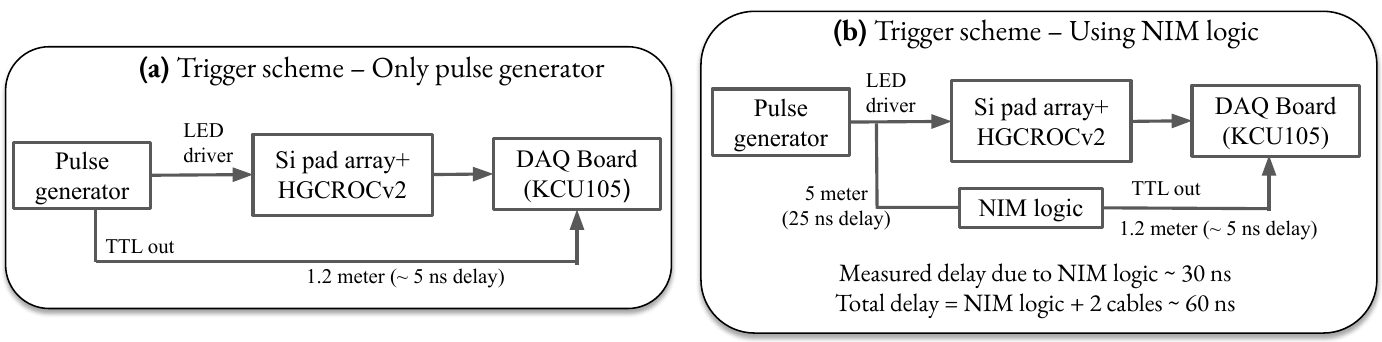}
\caption{The flow chart of the detector setup. (a) The pulse generator drives the LED and sends a trigger to the DAQ board. (b) The pulse generator drives the LED, with the trigger split into two paths: one directly to the DAQ board and the other via NIM logic to introduce an additional delay in the trigger.\label{fig:LED_pulserScheme}}
\end{figure}
This section reports the test performed on the packaged detectors using a blue LED light. The LED test is performed to configure the HGCROCv2 chip hosted by the FEE board. The HGCROCv2 is a highly integrated chip consisting of 72 readout channels with each channel consisting of a pre-amplifier, shaping amplifier, analog-to-digital converters (ADCs), and two time-to-digital converters (TDCs). The 10-bit ADC samples data at a 40~MHz clock with a selectable dynamic range of 80~fC, 160~fC, or 320~fC, depending on the gain setting of the pre-amplifier. When the ADC saturates, the two TDCs provide charge and time information, increasing the total dynamic range up to 10~pC. Sampled data is stored in a 10-bit (1024 samples) circular buffer, which releases the data upon receiving an external trigger. The data is then transmitted via two serial links (1.28 Gb/s each), one for data and the other for trigger information. The trigger phase (offset) is programmable and can be set in the HGCROCv2 configuration via the tuning of coarse and fine phase parameters. All the measurements with the LED and $^{90}$Sr reported here are done with the 80~fC dynamic range.\\
The HGCROCv2 was designed primarily for the Compact Muon Solenoid (CMS) experiment at the LHC, which plans to use p-type detectors. It can read both n-type and p-type Si pad detectors, however, charge injection for ASIC configuration is only possible with electrons (which are charge carriers in p-type detectors), not with holes (which correspond to the n-type detectors under investigation). Therefore to configure the chip for n-type detectors, an external injection using a blue LED light was performed by shining the LED directly onto the detector pads through the top-side opening on the FEE board (top panel in the Figure~\ref{fig:PadArrayAssembly} (d)). Configuration involves tuning the pedestal, adjusting pre-amplifier gain settings, and optimizing trigger parameters. Figure~\ref{fig:LED_pulser_photo} shows the LED test setup, which includes a pulse generator that provides peak-to-peak amplitude ($V_{pp}$) of 1.2~V with a 50\% duty cycle to drive the LED and provide a transistor-transistor logic (TTL) trigger signal to the Xilinx KCU105 AMD Kintex UltraScale FPGA KCU105 Evaluation Kit data acquisition (DAQ) board. The LED was shined through all the FEE openings to produce pulses in all 72 pads of the detector, as shown in Figure~\ref{fig:sr90-LED-pulses} (Left). The x-axis of the plot represents the ADC sampling clock, which samples every 25~ns~(40~MHz), and the y-axis shows the amplitude of the recorded pulse in ADC values. The figure shows that all 72 pads wire-bonded to the FEE board responded to the LED light, with their pulses peaking at 150~ns. The pulse amplitude during the LED scan varies from pad to pad because the light cannot be shined directly onto each pad due to the presence of transparent glue used as a glop top. Therefore, the LED test was performed at different angles through each opening to get a signal in each pad. This test aimed to check and confirm that all 72 pads were working and responding to the external injection.\\
The time at which the pulse reaches its maximum ADC value depends on the coarse and fine phase parameters configured in the chip, which vary with changes in the setup, such as additional delays caused by cables and NIM modules. To estimate the delay in phase due to the NIM modules, the signal from the pulse generator was split with one sending to drive the LED and the other through the NIM logic which consists of a leading edge discriminator and a  NIM to TTL converter module. The TTL signal compatible with the DAQ board is then sent to the DAQ board as shown in Figure~\ref{fig:LED_pulserScheme}. The acquired LED pulse showed a 60~ns delay, shifting the peak from 150~ns to around 90~ns. This confirms that the ADC timing is relative to the end of the buffer (512~bits in this case). Therefore it is important to adjust the trigger offset parameter accordingly. These parameters are kept constant during tests with both the LED and the $^{90}$Sr source to ensure a fair comparison of their pulses (see Figure~\ref{fig:sr90-LED-pulses}, right panel). More discussion on the detector testing with the $^{90}$Sr source is detailed in the following section.


\subsection{Detector test with $^{90}$Sr $\beta^{-}$ source }\label{subsec:sr90-test}

\begin{figure}[t]
\centering
\includegraphics[width=1.0\textwidth]{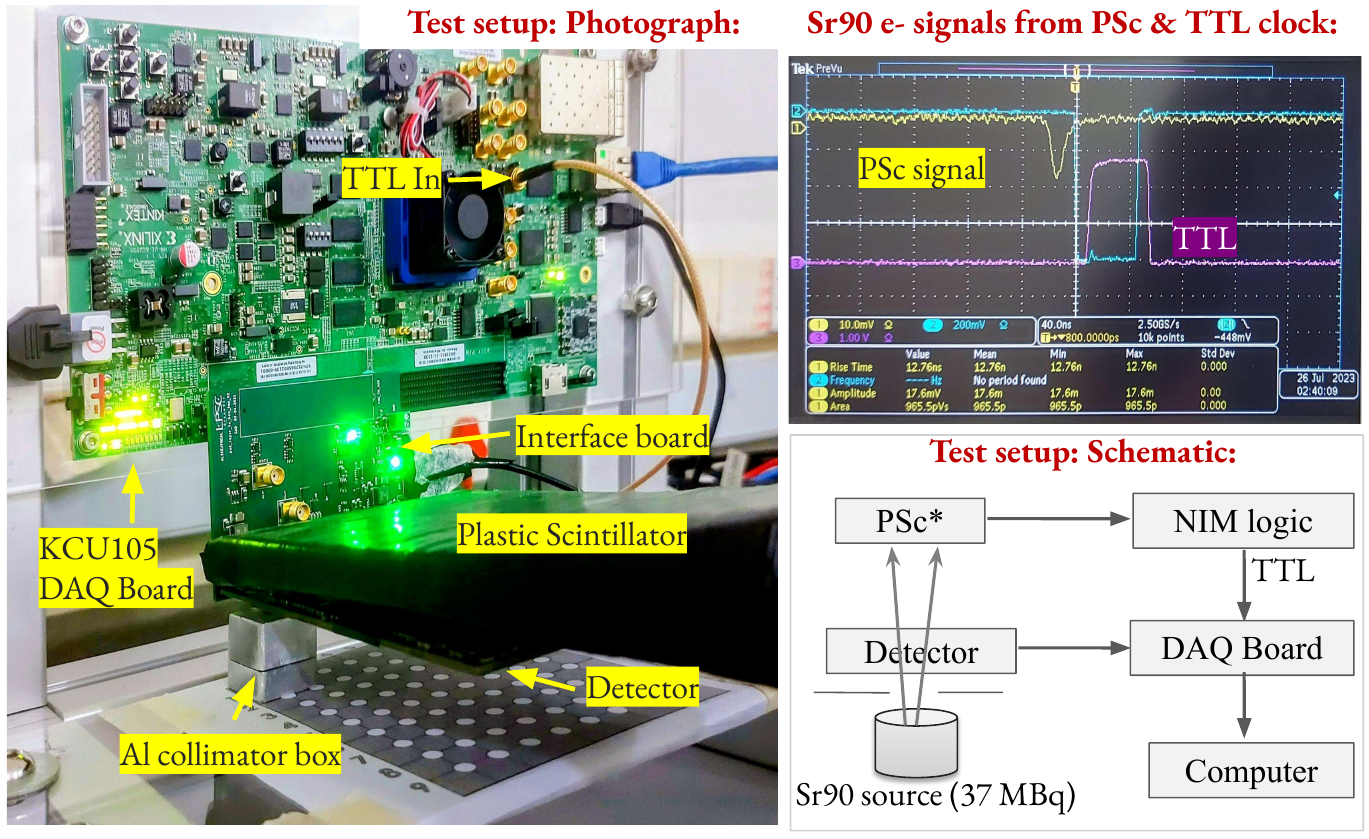}
\caption{(Left) Test setup with the Si detector assembly connected to the KCU105 DAQ board via an interface board, a $^{90}$Sr source inside an Al collimator, and a plastic scintillator above the detector. The scintillator sends a NIM signal to a discriminator, which converts it to TTL and sends it to the DAQ board. (Right-Top) NIM signal from the scintillator and TTL signal from the discriminator. (Right Bottom) Schematic of the test setup. \label{fig:sr90-SetupPhoto}}
\end{figure}

To test the detector with electrons, a 37~MBq $^{90}$Sr $\beta^{-}$ source was used, which produces electrons with energies up to 2.2~MeV. A dedicated test setup was created, as shown in Figure~\ref{fig:sr90-SetupPhoto}. The setup includes the detector connected to the DAQ board through an interface card. The collimated $^{90}$Sr source was placed inside a 6~mm thick aluminum collimator with a circular opening of 5~mm diameter to constrain the 2.2 MeV electrons onto one of the pad cells. The 6~mm thickness was selected based on the CSDA (continuous-slowing-down approximation) range, which indicates that 5~mm of aluminum fully absorbs electrons up to 2.5 MeV~\cite{NIST}. The detector was reverse-biased and kept at 60~V, while the scintillator placed above the detector was powered at 1200~V. The setup schematic is shown in Figure~\ref{fig:sr90-SetupPhoto} (Right-Bottom). The signal generated by the electrons in the scintillator, along with the TTL signal obtained from the NIM logic, is shown in Figure~\ref{fig:sr90-SetupPhoto} (Right-Top). The entire setup was operated in the dark to reduce the leakage current due to light. The right panel of Figure~\ref{fig:sr90-LED-pulses} shows the $^{90}$Sr electron pulse in comparison with the LED pulse. The $^{90}$Sr electron pulse has a much smaller amplitude compared to the LED light-induced signals. This is because the high-intensity LED light generates many more electron-hole pairs compared to a single electron passing through the 325-micron thick silicon, resulting in a larger pulse.\\

\begin{figure}[t]
\centering
\includegraphics[width=.4\textwidth]{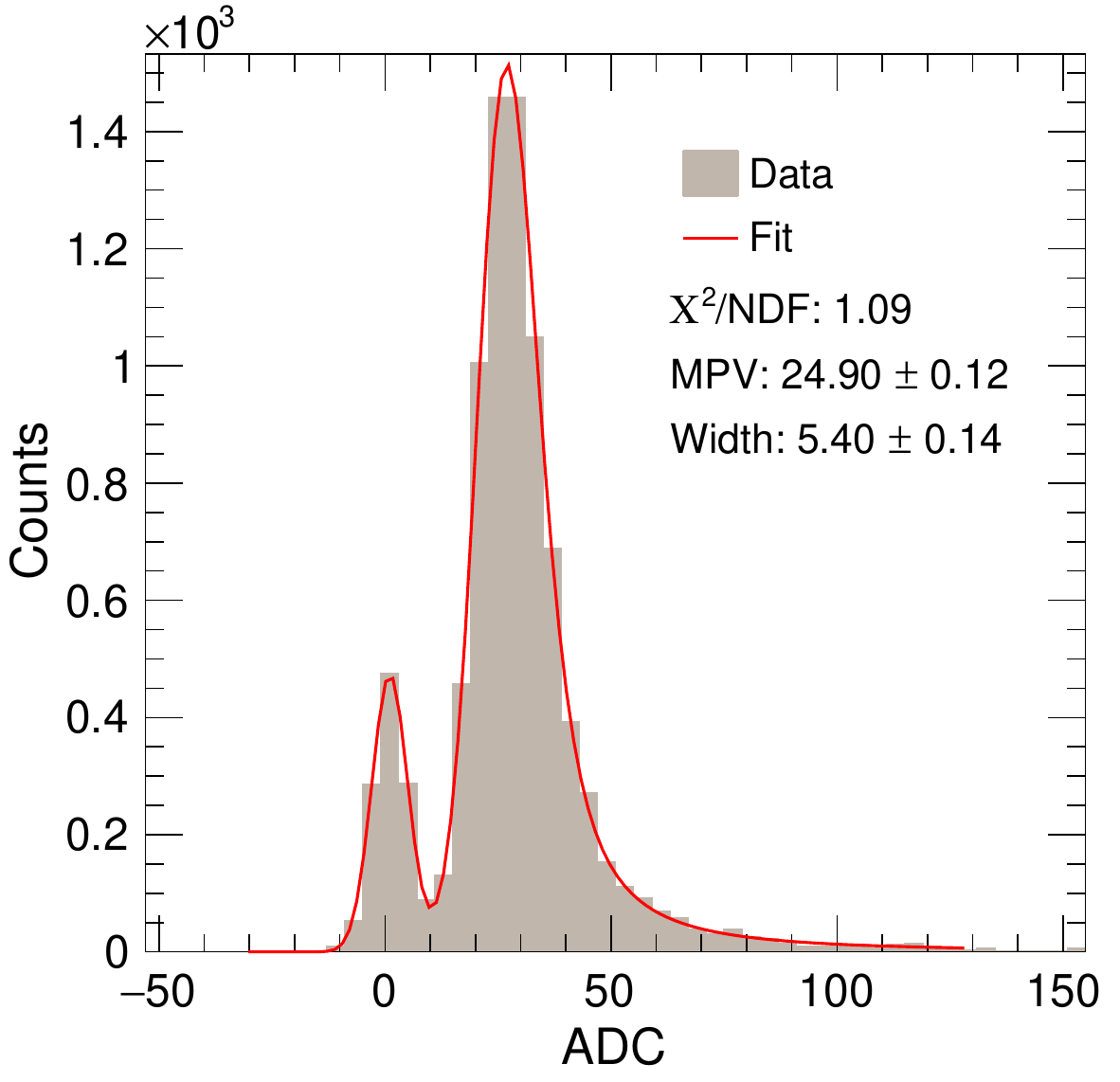}
\qquad
\includegraphics[width=.4\textwidth]{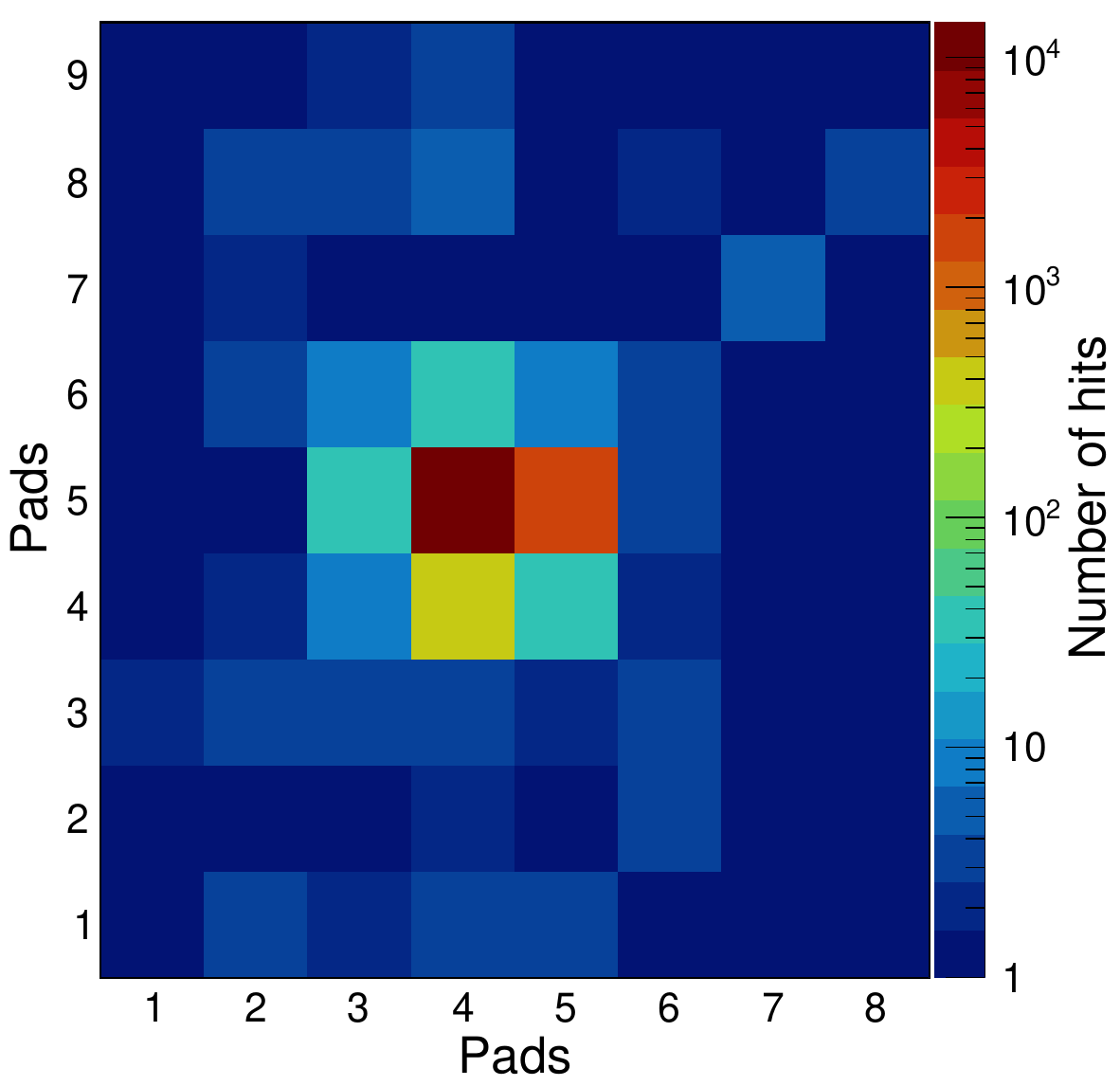}
\caption{(Left) Energy-loss distribution of electrons coming from $^{90}$Sr source. It is fitted with a Gaussian plus a convolution of Landau and Gaussian functions. The Gaussian fits the pedestal while the convoluted function fits the signal. (Right) Hitmap of $^{90}$Sr electrons placed below the detector.\label{fig:Sr90-MIP-fit_hitmap}}
\end{figure}
\begin{figure}[ht]
\centering
\includegraphics[width=.45\textwidth]{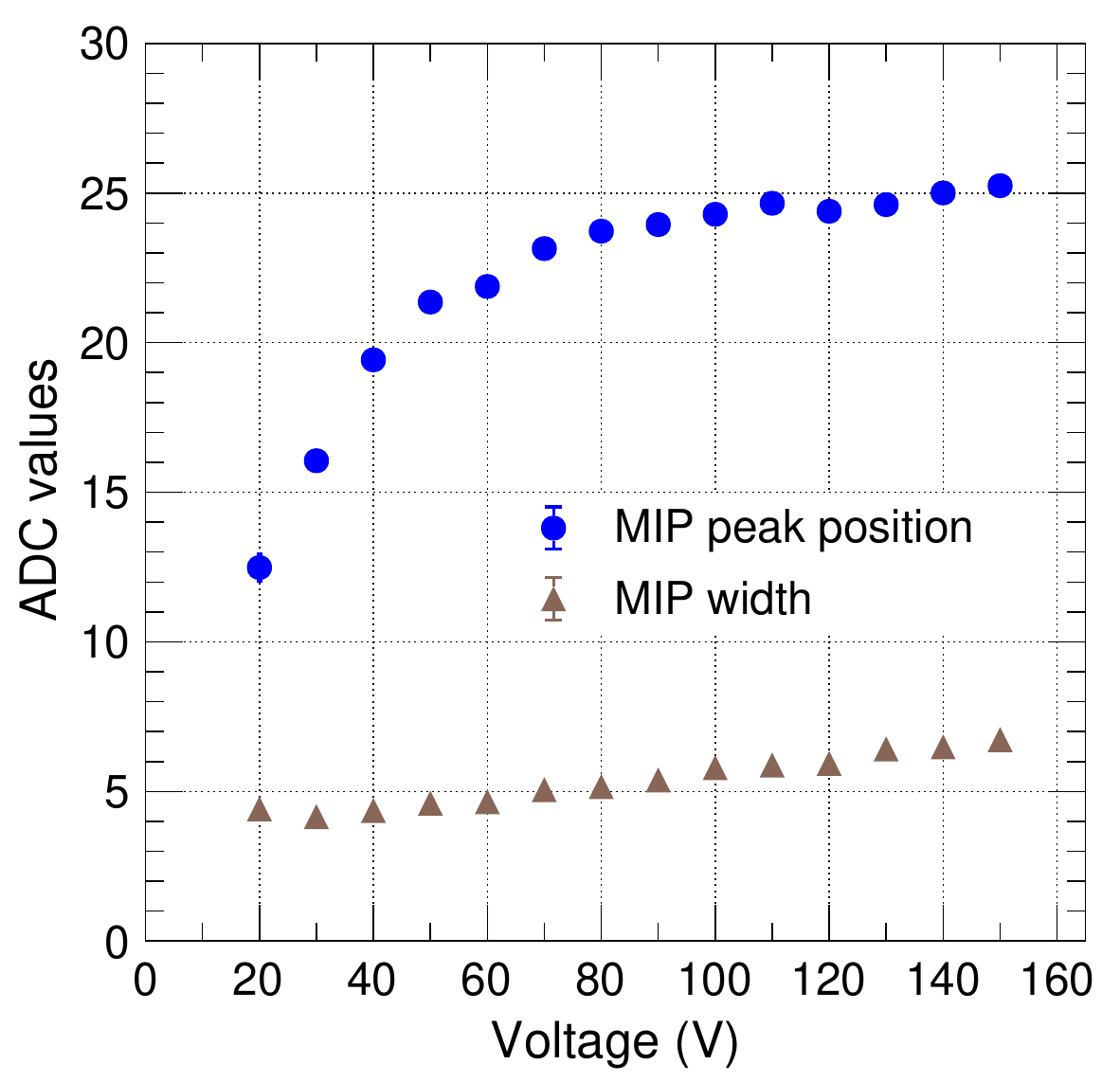}
\qquad
\includegraphics[width=.45\textwidth]{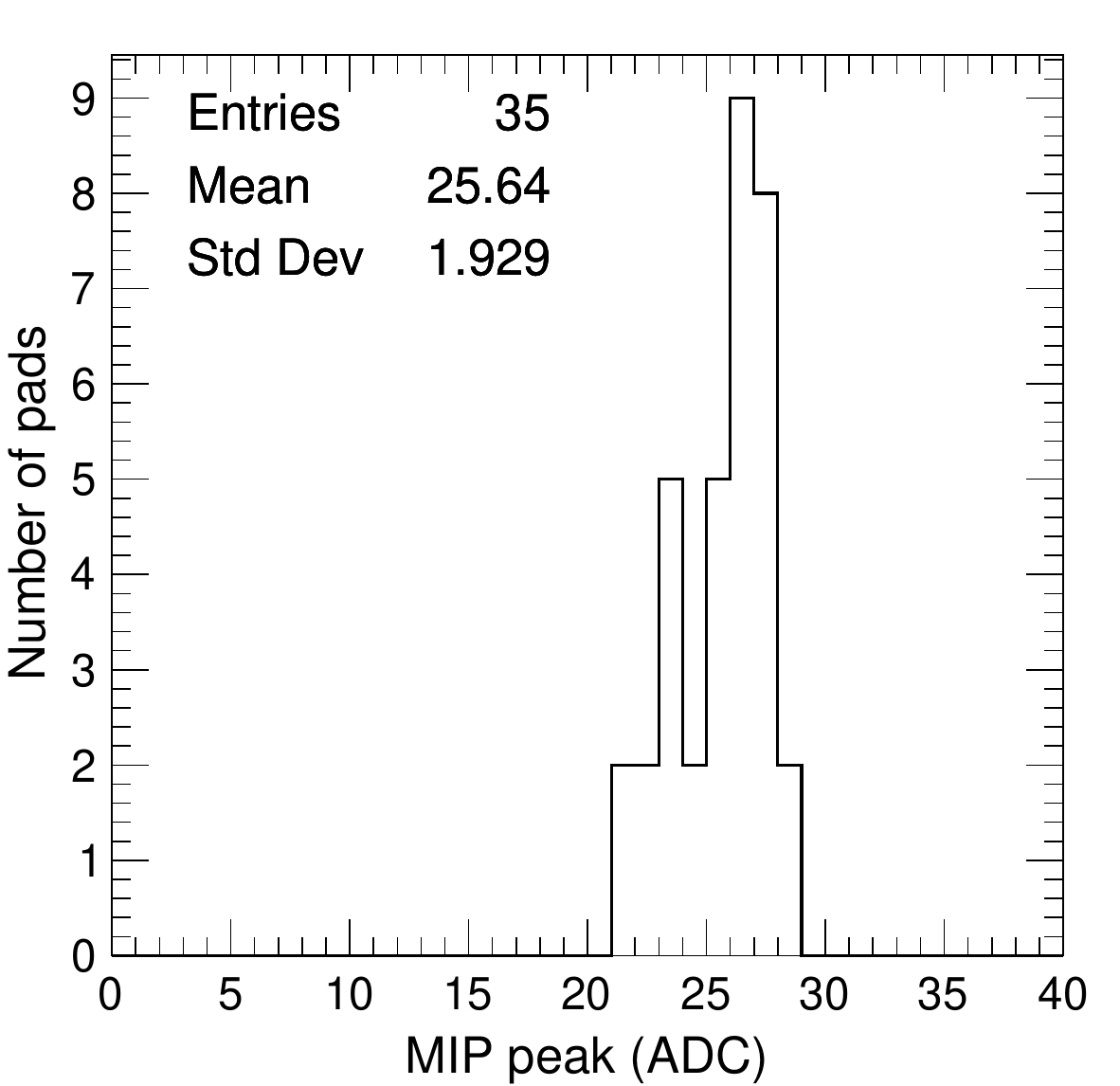}
\caption{(Left) $^{90}$Sr energy loss signal peak (ADC values) as a function of reverse bias voltage for one of the Si pad cells. The peak position shifts to higher values with an increase in the reverse bias voltage and then starts to saturate after full depletion occurs in the detector. (Right) Position scan using the electrons from $^{90}$Sr source, showing the homogeneity in the response of various detector pads selected randomly for the test. \label{fig:BiasVol-position_MIP}}
\end{figure}
 The measured energy-loss distribution of the $^{90}$Sr electrons in the detector is shown in Figure~\ref{fig:Sr90-MIP-fit_hitmap} (Left). The offset is removed, so the pedestal peaks at zero ADC value. The distribution is fitted with a Gaussian plus a convolution of Landau and Gaussian functions. The Gaussian fits the pedestal, while the Landau-Gaussian convolution fits the Minimum Ionizing Particle (MIP) peak. A MIP is a particle whose mean energy loss rate through matter is close to the minimum. Electrons with energy of few MeV acts approximately as MIP. Figure~\ref{fig:Sr90-MIP-fit_hitmap} (Right) shows the electron hit position on the detector array, indicating energy deposition in a few pad cells.

A voltage scan was conducted to measure the effect of reverse bias voltage on the electron energy-loss distribution. Figure~\ref{fig:BiasVol-position_MIP} (Left) shows the peak and width of the energy-loss distribution as functions of reverse bias voltage. The peak position increases with applied voltage and starts to saturate around 80~V, while the width remains mostly unchanged. The peak saturation occurs because the detector achieves full depletion at around 50~V.
Additionally, a position scan was also performed to measure the homogeneity in the response of different detector pads to energy loss distributions. Various pad cells were irradiated with the $^{90}$Sr source under the same conditions and their MIP peak position values were recorded as shown in Figure~\ref{fig:BiasVol-position_MIP} (Right). Most tested pads have a mean value of around 25~ADC values with a deviation of about 2~ADC value, indicating minor variations in pad response to electron energy loss.


\subsection{Radiation hardness studies}
\begin{figure}[t]
\includegraphics[width=0.5\linewidth]{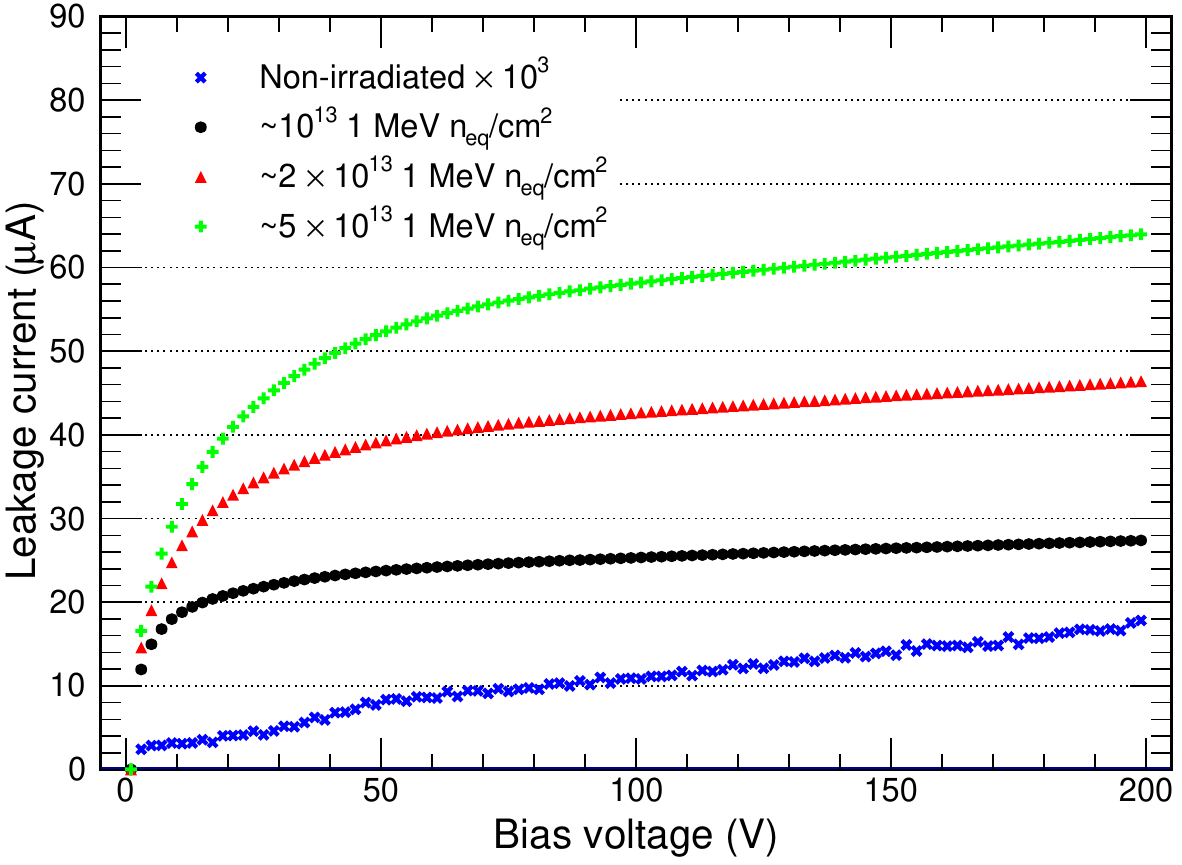}
\includegraphics[width=0.495\linewidth]{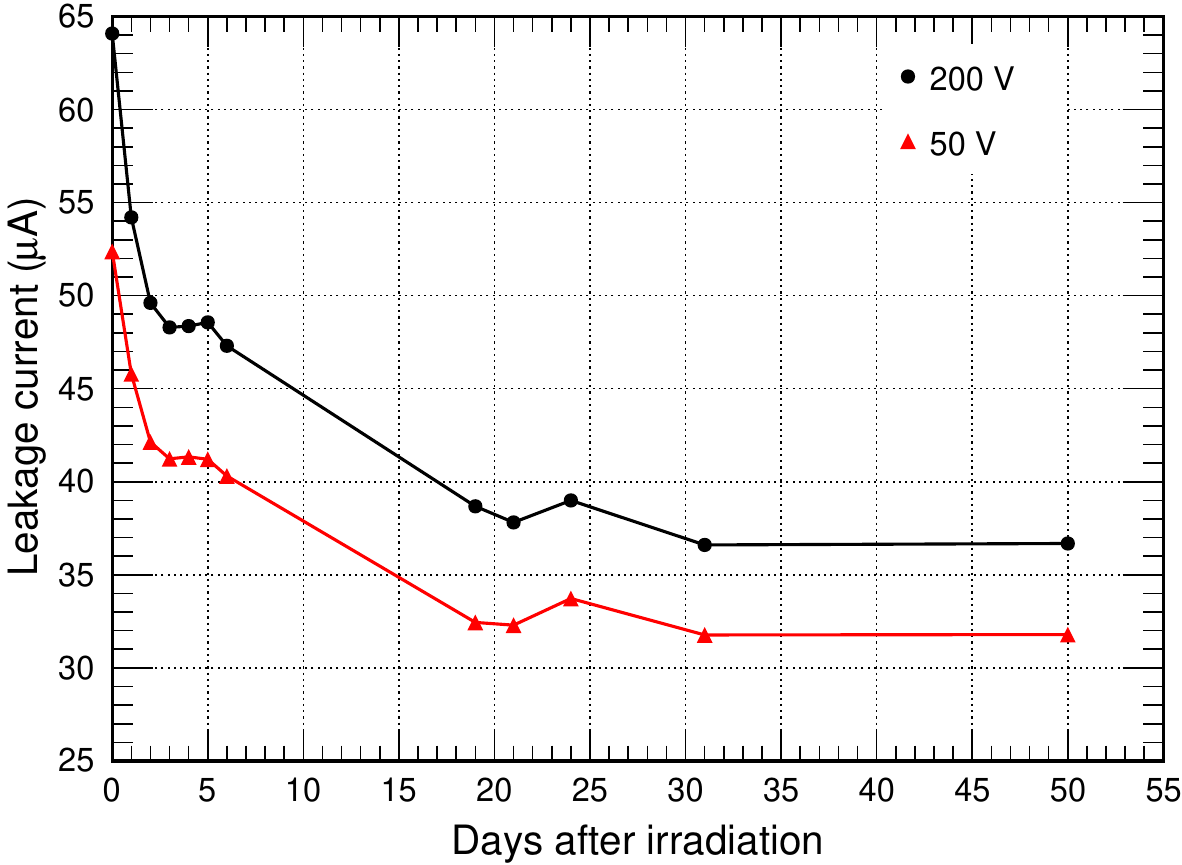}
\caption{(Left) IV curve comparison of non-irradiated and irradiated single pad cell as a function of radiation fluence. The non-irradiated leakage current data is scaled by~10$^{3}$. (Right) The leakage current was measured after irradiation for 55 days at two fixed reverse bias voltages.\label{fig:radHardIV}}
\end{figure}

During the LHC Run~4, spanning about four years, the innermost part of the ALICE FoCal detector will be exposed to the radiation equivalent to $7\times10^{12}$~1~MeV~$\mathrm{n_{eq}}$/cm$^{2}$. To test the performance of the fabricated detectors under these conditions, several test samples of $1\times 1$~cm$^{2}$ pads were irradiated with fast neutrons up to the fluence of $5\times10^{13}$~1~MeV~$\mathrm{n_{eq}}$/cm$^{2}$ at the RIKEN Accelerator-driven compact Neutron Systems (RANS) in Japan, where neutrons are produced using the proton-beryllium (p-Be) reaction~\cite{RANS_Facility}.
The leakage current of the irradiated pads was measured at different neutron fluence levels as a function of reverse bias voltage at room temperature, as shown in Figure~\ref{fig:radHardIV} (Left). As expected, the irradiated pad cell current increased with neutron fluence. The leakage current increased by three orders of magnitude for a fluence of $10^{13}$~1~MeV $\mathrm{n_{eq}}$/cm$^{2}$ compared to a non-irradiated test sample (blue data points scaled by $10^3$). Figure~\ref{fig:radHardIV} (Right) shows the current of the irradiated pad cell monitored over a period of 55 days after irradiation, where the current decreased from 64~$\muup$A to 47~$\muup$A within a week. The irradiated pads showed no response to the electron source, indicating that the detector might not tolerate the given fluence and the substrate-type inversion could have occurred~\cite{Moll:1999kv}. Therefore, using p-type substrate-based detectors is recommended to meet the high radiation tolerance requirements, which operate stably up to a fluence of $5\times10^{13}$ 1 MeV $\mathrm{n_{eq}}$/cm$^{2}$, provided they are maintained below 20\textdegree C~\cite{focaltdr}.
\section{Summary}
The 8$\times$9 n-type silicon pad array detectors were successfully designed and fabricated at Bharat Electronics Limited, India. The detectors met the design specifications for leakage current, full depletion voltage, and capacitance. They were packaged using a FEE board with HGCROCv2 ASIC, and the ASIC was configured by externally shining an LED onto the detector pads. The packaged detectors were tested with $^{90}$Sr electrons to perform voltage scans, position scans, and measure energy deposition in the detector's active volume. The voltage scan showed the variation of the $^{90}$Sr peak position with voltage, which saturated after reaching full depletion. The position scan demonstrated the homogeneity of response across different detector pads. The energy deposition in the detector showed a clear separation of the pedestal and electron signal. The radiation hardness tests revealed that the leakage current in the detector increased with a neutron fluence of $5\times10^{13}$~1 MeV~$\mathrm{n_{eq}}$/cm$^{2}$. Although the leakage current decreased significantly over two months post-irradiation, it remained high enough to cause weak signals like MIP to merge with the noise, indicating that the n-type silicon detectors are unsuitable for high-radiation environments. Therefore, p-type substrate-based detectors are recommended for the high radiation tolerance required by the ALICE FoCal project.
\acknowledgments
The authors would like to thank the ALICE-FoCal and ALICE-India collaboration for their constant support throughout the project work. They also thank Mr. Debasis Barik, Mr. Deepak Kumar (Scientific Assistants, CMRP NISER), and Mr. Samar Mohan Mohanty (Project associate, CMRP) for their constant support throughout the project work.  Additionally, the authors would like to thank DAE and DST India for their financial support through the project entitled "Indian participation in the ALICE experiment at CERN," and the work is also
partly funded through the J.C. Bose fellowship of DST awarded to BM.

\bibliographystyle{JHEP}
\bibliography{biblio.bib}

\providecommand{\href}[2]{#2}\begingroup\raggedright\begin{thebibliography}{10}

\bibitem{ATLASpixeldetector}
G.~Aad, M.~Ackers, F.~Alberti, M.~Aleppo, G.~Alimonti, J.~Alonso et~al., \emph{Atlas pixel detector electronics and sensors}, {\emph{Journal of Instrumentation} {\bfseries 3} (2008) P07007}.

\bibitem{CMSbarrelPixelDetector}
Y.~Allkofer, C.~Amsler, D.~Bortoletto, V.~Chiochia, L.~Cremaldi, S.~Cucciarelli et~al., \emph{Design and performance of the silicon sensors for the cms barrel pixel detector}, {\emph{Nuclear Instruments and Methods in Physics Research Section A: Accelerators, Spectrometers, Detectors and Associated Equipment} {\bfseries 584} (2008) 25}.

\bibitem{focal_LOI}
{\scshape ALICE} collaboration, \emph{{Letter of Intent: A Forward Calorimeter (FoCal) in the ALICE experiment}},  Tech. Rep. \href{https://cds.cern.ch/record/2719928}{CERN-LHCC-2020-009, LHCC-I-036}, CERN, Geneva (2020).

\bibitem{focaltdr}
{\scshape ALICE} collaboration, \emph{{Technical Design Report of the ALICE Forward Calorimeter (FoCal)}},  Tech. Rep. \href{https://cds.cern.ch/record/2890281}{CERN-LHCC-2024-004, ALICE-TDR-022}, CERN, Geneva (2024).

\bibitem{Clice_ILC_1}
{\scshape CALICE} collaboration, \emph{{Design and Electronics Commissioning of the Physics Prototype of a Si-W Electromagnetic Calorimeter for the International Linear Collider}}, \href{https://doi.org/10.1088/1748-0221/3/08/P08001}{\emph{JINST} {\bfseries 3} (2008) P08001} [\href{https://arxiv.org/abs/0805.4833}{{\ttfamily 0805.4833}}].

\bibitem{Clice_ILC_2}
C.~Adloff, Y.~Karyotakis, J.~Repond, J.~Yu, G.~Eigen, C.~Hawkes et~al., \emph{Response of the calice si-w electromagnetic calorimeter physics prototype to electrons}, \href{https://doi.org/https://doi.org/10.1016/j.nima.2009.07.026}{\emph{Nuclear Instruments and Methods in Physics Research Section A: Accelerators, Spectrometers, Detectors and Associated Equipment} {\bfseries 608} (2009) 372}.

\bibitem{Clice_ILC_3}
K.~Kawagoe, Y.~Miura, I.~Sekiya, T.~Suehara, T.~Yoshioka, S.~Bilokin et~al., \emph{Beam test performance of the highly granular siw-ecal technological prototype for the ilc}, \href{https://doi.org/https://doi.org/10.1016/j.nima.2019.162969}{\emph{Nuclear Instruments and Methods in Physics Research Section A: Accelerators, Spectrometers, Detectors and Associated Equipment} {\bfseries 950} (2020) 162969}.

\bibitem{Tsukuba_paper}
T.~Awes, C.~Britton, T.~Chujo, T.~Cormier, M.~Ericson, N.~Ezell et~al., \emph{Design and performance of a silicon–tungsten calorimeter prototype module and the associated readout}, \href{https://doi.org/https://doi.org/10.1016/j.nima.2020.164796}{\emph{Nuclear Instruments and Methods in Physics Research Section A: Accelerators, Spectrometers, Detectors and Associated Equipment} {\bfseries 988} (2021) 164796}.

\bibitem{sanjib_paper}
S.~Muhuri, S.~Mukhopadhyay, V.~Chandratre, T.~Nayak, S.K.~Saha, S.~Thakur et~al., \emph{Fabrication and beam test of a silicon-tungsten electromagnetic calorimeter}, \href{https://doi.org/10.1088/1748-0221/15/03/P03015}{\emph{Journal of Instrumentation} {\bfseries 15} (2020) P03015}.

\bibitem{HGCROCv2_paper}
D.~Thienpont and C.~de~La~Taille, \emph{Performance study of {HGCROC-v2}: the front-end electronics for the {CMS High Granularity Calorimeter}}, \href{https://doi.org/10.1088/1748-0221/15/04/C04055}{\emph{Journal of Instrumentation} {\bfseries 15} (2020) C04055}.

\bibitem{testbeampaper}
Sawan, M.~Bregant, J.L.~Bouly, O.~Bourrion, A.~van~den Brink, T.~Chujo et~al., \emph{Beam test of n-type silicon pad array detector at ps cern},  2024.

\bibitem{TCAD_simulations}
E.~Guichard and I.~Silvaco, \emph{Silvaco tcad},  Sep, 2022.
\newblock doi:10.21981/MZFR-HK34.

\bibitem{bourrion2023prototype}
O.~Bourrion, D.~Tourres, R.~Guernane, C.~Arata, J.-L.~Bouly and N.~Ponchant, \emph{Prototype electronics for the silicon pad layers of the future forward calorimeter (focal) of the alice experiment at the lhc}, {\emph{Journal of Instrumentation} {\bfseries 18} (2023) P04031}.

\bibitem{bel}
``{Bharat Electronics Ltd, Bangalore, India}.'' \url{http://www.bel-india.in}.

\bibitem{leobook1994}
W.R.~Leo, \emph{Techniques for nuclear and particle physics experiments: a how-to approach}, Springer Science \& Business Media (1994).

\bibitem{NIST}
M.~Berger, J.~Coursey and M.~Zucker, \emph{Estar, pstar, and astar: Computer programs for calculating stopping-power and range tables for electrons, protons, and helium ions (version 1.21)},  1999-01-01, 1999.

\bibitem{RANS_Facility}
T.~Kobayashi, S.~Ikeda, Y.~Otake, Y.~Ikeda and N.~Hayashizaki, \emph{Completion of a new accelerator-driven compact neutron source prototype rans-ii for on-site use}, \href{https://doi.org/https://doi.org/10.1016/j.nima.2021.165091}{\emph{Nuclear Instruments and Methods in Physics Research Section A: Accelerators, Spectrometers, Detectors and Associated Equipment} {\bfseries 994} (2021) 165091}.

\bibitem{Moll:1999kv}
M.~Moll, \emph{{Radiation damage in silicon particle detectors: Microscopic defects and macroscopic properties}}, Ph.D. thesis, Hamburg U., 1999.

\end{thebibliography}\endgroup

\end{document}